\documentclass[journal]{IEEEtran}
\usepackage{amsmath,amsfonts}
\usepackage{algorithmic}
\usepackage{array}
\usepackage[caption=false,font=footnotesize,labelfont=rm,textfont=rm]{subfig}

\usepackage{multirow}
\usepackage{textcomp}
\usepackage{stfloats}
\usepackage{graphicx}
\usepackage{url}
\usepackage{verbatim}
\usepackage{cite}
\usepackage{makecell}
\usepackage{array}
\usepackage{multirow}
\usepackage{graphbox}
\usepackage{endnotes}
\usepackage{color}
\usepackage[table]{xcolor}
\usepackage{array}
\usepackage{xcolor}
\usepackage{amsthm}
\usepackage{tabularx}

\newtheorem{theorem}{Theorem}

\theoremstyle{remark}
\def\BibTeX{{\rm B\kern-.05em{\sc i\kern-.025em b}\kern-.08em
    T\kern-.1667em\lower.7ex\hbox{E}\kern-.125emX}}
\usepackage{balance}
\begin{document}
\newcolumntype{L}[1]{>{\raggedright\arraybackslash}p{#1}}
\newcolumntype{C}[1]{>{\centering\arraybackslash}p{#1}}
\newcolumntype{R}[1]{>{\raggedleft\arraybackslash}p{#1}}

\title{Explainable Bayesian Recurrent Neural Smoother to Capture Global State Evolutionary Correlations}
\author{Shi Yan, Yan Liang$\dagger$, Huayu Zhang, Le Zheng,~\IEEEmembership{Senior Member,~IEEE}, Difan Zou, Binglu Wang,~\IEEEmembership{Member,~IEEE}
         \thanks{Shi Yan, Yan Liang, Huayu Zhang and Binglu Wang are with the School of Automation, Northwestern Polytechnical University, Xi'an 710072, China (e-mail: yanshi@mail.nwpu.edu.cn; liangyan@nwpu.edu.cn; zhanghuayu@mail.nwpu.edu.cn; wbl921129@gmail.com.)}
        \thanks{Le Zheng is with the Radar Research Laboratory, School of Information and Electronics, Beijing Institute of Technology, Beijing 100081, China (e-mail: le.zheng.cn@gmail.com).}
        \thanks{Difan Zou is with the Department of Computer Science, the University of Hong Kong, Pokfulam 999077, Hong Kong (e-mail: dzou@cs.hku.hk)}
        \thanks{$\dagger$Corresponding author: Yan~Liang.}
	\thanks{This work was supported by the National Natural Science Foundation of China under Grant 61873205.}}

\markboth{Journal of \LaTeX\ Class Files,~Vol.~18, No.~9, September~2020}%
{How to Use the IEEEtran \LaTeX \ Templates}

\maketitle
\begin{abstract}
Through integrating the evolutionary correlations across global states in the bidirectional recursion, an explainable Bayesian recurrent neural smoother (EBRNS) is proposed for offline data-assisted fixed-interval state smoothing. At first, the proposed model, containing global states in the evolutionary interval, is transformed into an equivalent model with bidirectional memory. This transformation incorporates crucial global state information with support for bi-directional recursive computation. For the transformed model, the joint state-memory-trend Bayesian filtering and smoothing frameworks are derived by introducing the bidirectional memory iteration mechanism and offline data into Bayesian estimation theory. The derived frameworks are implemented using the Gaussian approximation to ensure analytical properties and computational efficiency. Finally, the neural network modules within EBRNS and its two-stage training scheme are designed. Unlike most existing approaches that artificially combine deep learning and model-based estimation, the bidirectional recursion and internal gated structures of EBRNS are naturally derived from Bayesian estimation theory, explainably integrating prior model knowledge, online measurement, and offline data. Experiments on representative real-world datasets demonstrate that the high smoothing accuracy of EBRNS is accompanied by data efficiency and a lightweight parameter scale.
\end{abstract}
\section{Introduction}\IEEEPARstart{S}{tate} estimation in dynamic systems, which involves restoring the system state from noisy observations, is a critical field of research with extensive applications in target tracking \cite{hao_auto}, navigation \cite{hao_tsp,navigation_intro}, and signal processing \cite{TSP_NOM}. In scenarios where some delay and computational cost are permissible, Bayesian smoothing (BS) offers higher estimation accuracy than Bayesian filtering (BF) by performing recursive backward smoothing after forward filtering, thus utilizing all available observations throughout the evolutionary interval \cite{Simon_book_2017}. Classical Bayesian smoothers include the Rauch Tung-Striebel (RTS) smoother \cite{RTS_1965}, also known as the Kalman smoother (KS), for Gaussian linear systems; the Gaussian approximation smoother \cite{simon_smoothing_2010}, including the extended KS (EKS) \cite{EKS} and the unscented KS (UKS) \cite{UKS_ori}, for Gaussian nonlinear systems; and the particle smoother (PS) \cite{PS_ori} for non-Gaussian nonlinear systems. Generally, the optimality of these model-based (MB) smoothers relies on the accuracy of prior model knowledge, such as model structure and parameters, which is often challenging to obtain accurately in practice, particularly in non-cooperative scenarios.

For adaptive smoothing under model uncertainty, online model learning by using measurement data has been studied based on Bayesian estimation (BE) theory, with its representative methods including multi-model smoothing and joint estimation and identification (JEI). 
The multi-model smoothing uses a set of candidate models that switch according to a Markov chain, combining their smoothing posteriors for estimation \cite{IMM_Smooth_1995_TIT,IMM_Smooth_2000_AES,IMM_Smooth_2017AES}. Its estimation performance is still heavily dependent on the prior information, including the model set and transition probabilities.
For unknown model parameters, JEI iteratively performs joint parameter identification and state estimation \cite{liu2021based,liu2022maneuvering}, with its expectation maximization-based implementations explored in smoothing problems \cite{Auto_EM_2022,arxiv_EM_2024}. These methods are restricted to explicit parametric models, which may be mismatched and challenging to design practically. The variational Bayesian inference based adaptive smoother, another JEI method, addresses model mismatches by identifying second-order moments, but often requires hard-to-verify assumptions like conjugate distributions \cite{VBAKS_2015,VBAKS_Huang_2016,VBAKS_Huang_2020,VBAKS_2021}. The above methods mainly focus on online learning from measurement data, resulting in significant computational costs and overlooking the abundant state evolution information in offline data.

Deep learning (DL), with its powerful nonlinear fitting capabilities, can effectively capture state evolution information implicit in offline data. Recurrent neural networks (RNNs), like long short-term memory (LSTM) networks \cite{LSTM}, distill essential state evolution information through memory iterations. The bidirectional RNN (Bi-RNN) \cite{BIRNN} extends this capability by incorporating both forward and backward temporal correlations. The attention mechanism, notably in Transformers \cite{Transformer}, further optimizes global observation information usage. Mamba \cite{MAMBA}, with its selective state space model architecture, processes long sequences more efficiently. Despite their remarkable data-learning capabilities, as seen in large language models \cite{GPT, Llama}, these DL models require vast parameters and datasets, which can dilute the physical explainability of their parameters. This makes them less efficient for learning dynamic systems with well-defined physical properties. Consequently, while these temporal DL models have been applied to state estimation under model uncertainties \cite{pure_lstm1,pure_lstm2,Trans_filter1,Trans_filter2}, they often suffer from a lack of explainability and demand extensive training data, even for simple sequence models.

To this end, combining MB methods with DL presents a feasible solution \cite{model_based_DL}, expected to enhance explainability and reduce data dependency by leveraging model knowledge, as well as enable efficient learning with fewer parameters. Existing MB and DL combinations mainly focus on filtering, with primary implementations including output compensation, model learning, and gain learning. The output compensation method employs the temporal DL model to compensate for MB filter estimates in a sequence-to-sequence format \cite{DeepMTT}. Its effectiveness depends on the accuracy of the base filter or smoother, and not correcting for the covariance causes it to be inconsistent with the corrected state. Model learning methods embed RNNs into MB filters as output mappings for state transitions and covariances, but this neglects prior model knowledge \cite{GRU_pf,CNN_pf,jung2020mnemonic,yan_arm}. In tasks about computer vision, observation models are also constructed by DL models to extract the high-dimensional observation features \cite{ICCV2017,TNNLS_Dynanet}. The gain learning method estimates the filtering gain using GRUs to balance prediction and measurement in filtering \cite{KalmanNet}. This method is further extended to a smoothing algorithm with bidirectional structures, estimating forward and backward gains separately with GRUs \cite{RTSNet}. By omitting the second-order moments, the gain-learning approach speeds up the computation, but consequently fails to utilize the prior noise statistics and measure the estimation covariance.

In fact, deriving an algorithmic framework grounded in well-defined models and BE theory to integrate offline data with prior model knowledge is crucial for enhancing estimation performance and explainability. In our previous work, an explainable Bayesian gated RNN for non-Markov state estimation has been developed by introducing DL into BF through rigorous mathematical derivations \cite{yan_tsp}. Its explainable computational architecture naturally integrates prior model knowledge and offline data, enabling effective model learning while maintaining a lightweight parameter scale and low training data dependency. This finding motivates our development of state smoothing, which implies performing explainable learning on complete sequential data that contains global state information. Therefore, it is required to derive data-assisted smoothing algorithms based on BE theory to capture the global state correlations over the evolutionary interval.

This paper presents a novel model for state smoothing in dynamic systems, considering the influence of all states within the evolutionary interval, excluding the current state, on the current evolution. For this model, an explainable Bayesian recurrent neural smoother (EBRNS) specific to the fixed-interval smoothing problem is derived. At first, the proposed high-dimensional model with global state information is transformed into an equivalent one with bidirectional memories, integrating global state evolutionary information while supporting the recursive processes of BF and BS. Secondly, the data-assisted joint state-memory-trend (JSMT) BF and JSMT-BS frameworks are derived based on the transformed model. A bidirectional filtering and smoothing structure consists of JSMT-BF and JSMT-BS, capturing the evolutionary correlations of global states by utilizing the bidirectional memory iteration in a well-defined gated structure. The Gaussian approximation implementation of the framework is derived by considering the computational efficiency and analytical properties. The bidirectional recursive architecture and internal gated structure of EBRNS are not artificial combinations but naturally derived from BE theory, which provide it with good explainability and optimality. Experiments with representative real-world datasets show that EBRNS delivers high smoothing accuracy, data efficiency, and a lightweight parameter scale.

The following notations are used throughout this paper: the conditional expectation is denoted by ${\rm{E}}\left[ { \cdot | \cdot } \right]$, and the squared two-norm operation is denoted by $\left \| \cdot \right \| ^{2}$; ${\rm{diag}}\left(d_1, d_2\right)$ represents a diagonal matrix with $d_1$ and $d_2$ as its diagonal elements; the notation $(\cdot)$ signifies the same content as that in the previous parentheses; the superscripts $(\cdot)^\top$ and $(\cdot)^{-1}$ indicate the transpose and inverse operations, respectively.

\section{Problem Formulation}
The state estimation of the dynamic system typically involves the use of the first-order Markov state space model with additive noise, i.e., 
\begin{flalign}
&\noindent \emph{System model:}&\notag \\
&\quad \quad \quad \quad \quad \quad \quad  {{\bf{x}}_k} = f_k\left( {{{\bf{x}}_{k - 1}}} \right) + {\bf{w}}_{k}&
\label{system_model}\\
&\noindent \emph{Measurement model:}&\notag \\
&\quad \quad \quad \quad \quad \quad \quad  {{\mathbf{z}}_k} = {h_k}\left({{\mathbf{x}}_k} \right) + {{\mathbf{v}}_k}&
\label{measurement_model}
\end{flalign}
where ${\bf x}_k$ and ${\bf z}_k$ indicate the state and measurement vectors, respectively, and their time index is $k$; the linear/nonlinear state-evolution function $f_k$ and state-measurement function $h_k$ are typically constructed a priori; the mutually independent ${\bf{w}}_{k}$ and ${{\bf{v}}_k}$ are process noise and measurement noise, respectively, and they always follow the Gaussian distribution. The dynamic model in Eq. (\ref{system_model}) is easy to construct, and its first-order Markov property enables recursive computation, making it widely used in various BF and BS algorithms. However, it overlooks the continuous effects of sequential states that are often present in practical state estimation tasks. For this problem, our previous work constructed a non-Markov state evolution model for filtering that considers the effect of the historical state on the current state evolution \cite{yan_tsp}, i.e.,
\begin{align}
{{\bf{x}}_k} = f_k^{\rm NM}\left( {{{\bf{x}}_{k - 1}},{{\bf{x}}_{k - 2}}, \cdots ,{{\bf{x}}_1}} \right)+{\bf w}_k
\label{no_markov_filtering_model}
\end{align}
where $f_k^{{\rm NM}}$ incorporates the effects of historical states, and Eq. (\ref{system_model}) is a special case of Eq. (\ref{no_markov_filtering_model}) if the states in $f_k^{\rm NM}$ include exclusively the previous one. Such a model is difficult to construct explicitly due to the increasing dimensionality of the state sequence over time. In \cite{yan_tsp}, it is transformed into an equivalent memory-based first-order Markov model that integrates historical state information while supporting efficient recursive computation. 
\begin{figure}[t]
\centering
\includegraphics[width=0.75\linewidth]{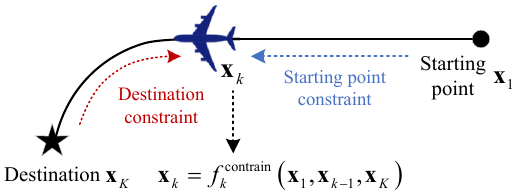}
\caption{Constraints from the starting point and the destination both have continuous effects on the state evolution of the target.}
\label{problem_example1}
\end{figure}

For fixed-interval smoothing, utilizing data across the entire time interval, Eq. (\ref{no_markov_filtering_model}) still falls short in information utilization, as current state evolution can be influenced by both historical and future states. Fig. \ref{problem_example1} illustrates a target tracking task using a constrained dynamic model, where constraints from the start point and destination continuously impact the model $f_k^{\rm constrain}$, and introducing these constraints enables to improve the modeling accuracy \cite{xu_destination}. Similar correlations are found in other time series fields: in text translation, understanding a word requires context; in weather forecasting, analysis needs to account for periodicity, seasonality, and trends. Thus, a new dynamic model for fixed-interval smoothing is proposed, considering the global state:
\begin{align}
 {{\bf{x}}_k} = {F_k}\left( {{{\bf{X}}_{\neg k}}} \right)  + {{\bf{w}}_k}
\label{smooth_model}
\end{align}
where ${{\bf{X}}_{\neg k}}=\{{{{\bf{x}}}_1}, \cdots ,{{\bf{x}}_{k - 1}},{{{\bf{x}}_{k + 1}}, \cdots ,{{\bf{x}}_K}}\}$, and $K$ is the duration time of the evolutionary interval. The function $F_k$ captures the effect of other $K-1$ states on the current state, and its time-varying and high-dimensional properties make it also difficult to construct a priori.

Furthermore, available offline data is considered, which consists of the ground-truth state sequences and their corresponding measurements, namely, 
\begin{flalign}
{\cal D} = {\left\{ {{\bf{x}}_{1:K}^i,{\bf{z}}_{1:K}^i} \right\}_{i = 1 \cdots I}}&
\label{offline_data}
\end{flalign}
where $I$ is the total number of samples, ${\bf{x}}_{1:K}^i$ and ${\bf{z}}_{1:K}^i$ are the ground-truth state and measurement sequences, respectively. Such offline datasets can be constructed by recording the noisy readings of sensors or performing simulations. For example, recording the flight states of an aircraft and simulating the corresponding radar observations.

Our goal is to naturally integrate DL and BE theory through rigorous mathematical derivations to deal with the fixed-interval smoothing problem under model uncertainty. Despite our successful experience in filtering \cite{yan_tsp}, this is still a challenging open issue. On the one hand, the model in Eq. (\ref{smooth_model}) needs to be transformed to integrate global state correlations while supporting recursive computation. On the other hand, an offline data-assisted BS method needs to be derived to reasonably integrate the prior model knowledge and offline data in an explainable architecture.

\section{Explainable Bayesian Recurrent Neural Smoother}
This section develops the EBRNS for fixed-interval smoothing. At first, the dynamic model in Eq.(\ref{smooth_model}) is transformed into an equivalent one with the bidirectional memory. After that, the offline data-assisted JSMT-BF and JSMT-BS are derived for the transformed model, resulting in the bidirectional recursive process and internal gated structure of EBRNS. The proposed frameworks are then implemented using Gaussian approximation. 
Finally, we design the neural network (NN) modules within EBRNS and its training scheme.

\subsection{Problem transformation}
\label{Method_A}
There are generally two operational stages of BS for fixed-interval smoothing, i.e., forward filtering and backward smoothing. In forward filtering, the state is iteratively filtered using real-time measurements. After that, an iterative refinement of the filtered state is performed in backward smoothing using all measurements in the evolutionary interval \cite{Simon_book_2017}. A similar bidirectional architecture is employed in Bi-RNN to capture and synthesize the forward and backward evolutionary information of states over time. Motivated by this, the model in Eq. (\ref{smooth_model}) is transformed into a two-stage model containing both forward and global trends, enabling it to support forward-backward recursive estimation in a BS framework while integrating global state information. 

At first, $f_k$ in Eq. (\ref{system_model}) and $f_k^{\rm NM}$ in Eq. (\ref{no_markov_filtering_model}) are substituted into Eq. (\ref{smooth_model}), such that the state estimation problem with considering global state information is transformed into the learning of corresponding evolution trend terms under a nominal first-order Markov model, namely,
\begin{align}
{{\bf{x}}_k} &= {f_k}\left( {{{\bf{x}}_{k-1}}} \right) + \underbrace {f_k^{\rm NM}\left( {{{{\bf{x}}_1}, \cdots ,{{\bf{x}}_{k - 1}}}} \right) - {f_k}\left( {{{\bf{x}}_{k-1}}} \right)}_{{\bf \Delta} _k^a} \notag \\&+ \underbrace {{F_k}\left( {{{\bf{X}}_{\neg k}}} \right) - f_k^{\rm NM}\left( {{{{\bf{x}}_1}, \cdots ,{{\bf{x}}_{k - 1}}}} \right)}_{{\bf \Delta} _k^b} + {{\bf{w}}_k}
\notag \\
&= {f_k}\left( {{{\bf{x}}_{k-1}}} \right) + {\bf \Delta} _k^a + {\bf \Delta} _k^b + {{\bf{w}}_k}
\label{smooth_trans1}
\end{align}
where ${\bf \Delta} _k^a$ is the forward trend that characterizes the effects of historical states on the current state evolution; ${\bf \Delta} _k^b$ is the global trend that characterizes the effects of the other $K-1$ states on the current state evolution. This two-stage process aims to simplify the development of both the filtering and subsequent smoothing frameworks. As ${\bf \Delta} _k^b$ contains all the information of the $K-1$ states, such a transformation is informationally equivalent, with only the addition of the processing stage. Given that non-Markov ${\bf \Delta} _k^a$ and ${\bf \Delta} _k^b$ lack support for recursive computation, we handle them using the concept of function nesting.

\subsubsection{Processing of the forward trend}
In the forward filtering stage, ${\bf \Delta} _k^a$ is correlated with historical states and approximated by a two-layer nested function, namely, 
\begin{align}
{\bf \Delta} _k^a = f_k^a( {\underbrace {g_k^a\left( {{{\bf{x}}_1}, \cdots ,{{\bf{x}}_{k - 1}}} \right)}_{{\bf{c}}_k^a}} )
\label{f_error}
\end{align}
where the function $g_k^a$ condenses the regularity information in historical states into the forward memory ${{\bf{c}}_k^a}$; The output function $f_k^a$ maps ${{\bf{c}}_k^a}$ to the forward trend. To support forward recursive filtering, we developed an iterative update process for ${{\bf{c}}}_k^a$, inspired by the memory update mechanism in LSTM \cite{LSTM}, namely,
\begin{align}
{\bf{c}}_k^a = {g^{\rm forward}}\left( {{\bf{c}}_{k - 1}^a,{{\bf{x}}_{k - 1}}} \right)
\label{f_mem_iteration}
\end{align}
where, at each frame, ${g^{\rm forward}}$ selectively retains crucial information in ${\bf{c}}_{k - 1}^a$ and introduces key information in ${\bf{x}}_{k - 1}$ to obtain ${\bf{c}}_{k}^a$. 

\subsubsection{Processing of the global trend}
As ${\bf \Delta} _k^b$ involves state information after time $k$, it is handled in the backward smoothing stage using a nested computation akin to that of ${\bf \Delta} _k^a$, namely,
\begin{align}
{\bf \Delta} _k^b = f_k^b( {\underbrace {g_k^b\left( {{{\bf{X}}_{\neg k}}} \right)}_{{\bf{c}}_k^b}})
\label{b_error}
\end{align}
where the function $g_k^b$ condenses the effects of the other $K-1$ states on the current state evolution, and $f_k^b$ is an output function. The learning of ${\bf \Delta} _k^b$ takes place in the backward smoothing recursion, enabling the consideration of backward state correlations. Correspondingly, ${{\bf{c}}_k^b}$ obtained by its condensation is named backward memory. The iterative process of ${\bf{c}}_K^b$ is described as follows. Given that the ${\bf{c}}_K^a$ obtained after the filtering stage contains the forward information for all $K$ states, it is used to initialize ${\bf{c}}_K^b$, i.e., ${\bf{c}}_K^b = {\bf{c}}_K^a$. After that, ${\bf{c}}_k^b$ is iterated backward over time to capture the backward correlations of states, namely, 
\begin{align}
{\bf{c}}_k^b = {g^{\rm backward}}\left( {{\bf{c}}_{k + 1}^b,{{\bf{x}}_{k + 1}}} \right)
\label{b_mem_iteration}
\end{align}
where ${g^{\rm backward}}$ introduces important information from ${\bf{x}}_{k + 1}$ and selectively retains information from ${\bf{c}}_{k + 1}^b$ at each frame.

\emph{Remark 1: The evolution correlations in the state sequences are condensed by $g_k^a$ and $ g_k^b $ into low-dimensional hidden variables $ \mathbf{c}_k^a $ and $ \mathbf{c}_k^b $.  This state-driven condensation process, similar to human memory mechanisms, requires learning $ g_k^a $ and $ g_k^b $ from offline data.}

Through the above transformation, the model considering global state in Eq. (\ref{smooth_model}) is equivalently converted to a model with the bidirectional memory iterative mechanism in Eqs. (\ref{smooth_trans1})-(\ref{b_mem_iteration}). The transformed model supports recursive filtering and smoothing computations while incorporating global state evolution correlations. Given offline data, the learning of ${\bf \Delta}^a_k$ and ${\bf \Delta}^b_k$ is transformed into the learning of $f_k^a$, $g^{\rm forward}$, $f_k^b$, and $g^{\rm backward}$. A corresponding Bayesian recursive framework is required to be derived to effectively integrate offline data, online measurements, and prior model knowledge through explainable learning.

\subsection{Explainable Bayesian gated framework}
Through introducing the offline dataset $\cal D$ in BF, the JSMT-BF is shown in Theorem \ref{Lem}.

\begin{theorem}
\label{Lem}{(Joint state-memory-trend Bayesian filtering)}
For the measurement model in Eq. (\ref{measurement_model}) and the dynamic model consisted by Eqs. (\ref{smooth_trans1})-(\ref{b_mem_iteration}), given the offline data set $\mathcal{D}$, the previous measurement sequence ${{\mathbf{z}}_{1:k - 1}}$, and the previous JSMT filtering posterior PDF $p\left( {{{\mathbf{x}}_{k - 1}},{{\mathbf{c}}_{k - 1}^a}\left| {{{\mathbf{z}}_{1:k - 1}},\mathcal{D}} \right.} \right)$, the JSMT filtering PDF prediction is
\begin{align}
&p\left( {{{\mathbf{x}}_k},{{\mathbf{c}}_k^a}\left| {{{\mathbf{z}}_{1:k - 1}}}{,\mathcal{D}} \right.} \right) 
\notag \\
&= \int\!\!\!\!\int {P_k^1p\left( {{{\mathbf{x}}_{k - 1}},{{\mathbf{c}}_{k - 1}^a}\left| {{{\mathbf{z}}_{1:k - 1}},\mathcal{D}} \right.} \right)}d{{\mathbf{x}}_{k - 1}}d{{\mathbf{c}}_{k - 1}^a}
\label{T1}
\end{align}
with
\begin{align}
P_k^1 =& \int {p\left( {{{\mathbf{x}}_k}\left| {{{\mathbf{x}}_{k - 1}},{{\bf \Delta} _k^a}} \right.} \right)P_k^2d {{\bf \Delta} _k^a}}
\label{T2}\\
P_k^2 =& p\left( {{\bf \Delta} _k^a\left| {{{\mathbf{c}}_k^a},\mathcal{D}} \right.} \right)p\left( {{{\mathbf{c}}_k^a}\left| {{{\mathbf{x}}_{k - 1}},{{\mathbf{c}}_{k - 1}^a},\mathcal{D}} \right.} \right)
\label{T3}
\end{align}
Given the current measurement ${\bf z}_k$, the JSMT filtering PDF update is
\begin{align}
p( {{{\mathbf{x}}_k},{{\mathbf{c}}_k^a}\left| {{{\mathbf{z}}_{1:k}},\mathcal{D}} \right.} )  \propto { {p\left( {{{\mathbf{z}}_k}\left| {{{\mathbf{x}}_k}} \right.} \right)p\left( {{{\mathbf{x}}_k},{{\mathbf{c}}_k^a}\left| {{{\mathbf{z}}_{1:k - 1}},\mathcal{D}} \right.} \right)} }
\label{T4}
\end{align}
\label{Bayesian_RNN}
\end{theorem}
\begin{proof}
See the proof of Theorem 1 in our previous work \cite{yan_tsp}, where the compensation for the observation model is not considered.
\end{proof} 

According to the two-stage processing, only ${\bf \Delta}_k^a$ in Eq. (\ref{smooth_trans1}) is compensated in the forward filtering stage, with ${\bf \Delta}_k^b$ addressed in the subsequent smoothing stage. Building on Theorem \ref{Lem}, we derive the JSMT-BS as presented in Theorem \ref{T}.
\begin{figure*}[t]
\centering
\includegraphics[width=1.\linewidth]{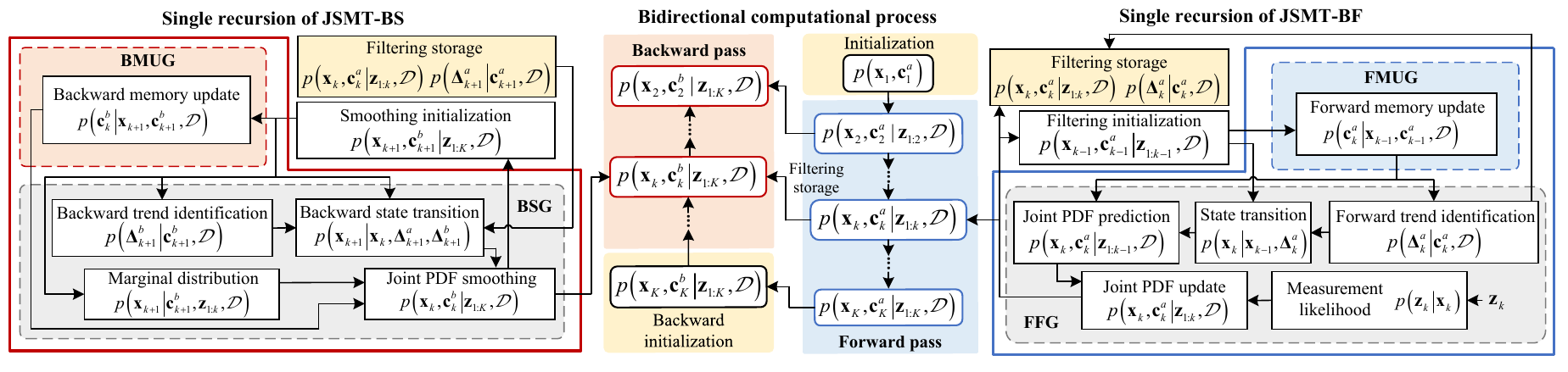}
\caption{The bidirectional transfer process is built through the collaboration of JSMT-BF and JSMT-BS.}
\label{prob_framework}
\end{figure*}

\begin{theorem}\label{T}{(Joint state-memory-trend Bayesian smoothing)}
For the measurement model in Eq. (\ref{measurement_model}) and the dynamic model in Eqs. (\ref{smooth_trans1})-(\ref{b_mem_iteration}), given the complete measurement sequence ${{\mathbf{z}}_{1:K}}$, the offline data set $\mathcal{D}$, the previous JSMT smoothing posterior density $p( {{{\bf{x}}_{k+1}},{\bf{c}}_{k+1}^b\left| {{{\bf{z}}_{1:K}},{\cal D}} \right.} )$, and the JSMT filtering posterior density $p( {{{\bf{x}}_k},{\bf{c}}_{k}^a\left| {{{\bf{z}}_{1:k}},{\cal D}} \right.} )$, the JSMT smoothing density is
\begin{align}
& p\!\left( {{{\bf{x}}_k},{\bf{c}}_k^b\left| {{{\bf{z}}_{1:K},{\cal D}}}\! \right.} \right) 
\!= \!\!\!\int\!\!\!\!\int\!\!\! {P_k^3} p( {{{\bf{x}}_{k + 1}},{\bf{c}}_{k + 1}^b\left| {{\bf{z}}_{1:K},{\cal D}} \right.} )d{{\bf{x}}_{k + 1}}d{\bf{c}}_{k + 1}^b
\label{smoothing1}
\end{align}
with
\begin{align}
&P_k^3 = {{p( {{\bf{c}}_k^b\left| {{{\bf{x}}_{k + 1}},{\bf{c}}_{k + 1}^b,{\cal D}} \right.} )P_k^4} \over {p( {{{\bf{x}}_{k + 1}}\left| {{\bf{c}}_{k + 1}^b,{{\bf{z}}_{1:k}},{\cal D}} \right.} )}}
\label{smoothing2}
\\
&P_{k + 1}^4 \!\! =\!\!\int\!\!\!\!\int\!\!\!\!\int\!\!\!\!\int\!\! {P_{k + 1}^5 }p\left( {{{\bf{x}}_k},{\bf{c}}_k^a\left| {{{\bf{z}}_{1:k}},{\cal D}} \right.} \right) d{\bf \Delta} _{k + 1}^ad{\bf \Delta} _{k + 1}^bd{\bf{c}}_{k+1}^ad{\bf{c}}_k^a
\\
&P_{k + 1}^5  = p( {{{\bf{x}}_{k + 1}}\left| {{{\bf{x}}_k},{\bf \Delta} _{k + 1}^a,{\bf \Delta} _{k + 1}^b} \right.} )p\left( {{\bf \Delta} _{k + 1}^a\left| {{\bf{c}}_{k + 1}^a,{\cal D}} \right.} \right)
\notag \\
&\quad \quad \quad \times 
p\left( {{\bf \Delta} _{k + 1}^b\left| {{\bf{c}}_{k + 1}^b,{\cal D}} \right.} \right) p\left( {{\bf{c}}_{k + 1}^a\left| {{\bf{c}}_k^a,{{\bf{x}}_k},{\cal D}} \right.} \right)
\end{align}
\label{Bayesian_Smoothing}
\end{theorem}
\begin{proof}
See Appendix \ref{Proof_A}.
\end{proof} 

Fig. \ref{prob_framework} shows the probabilistic transfer of the EBRNS, where JSMT-BF and JSMT-BS cooperatively form its bidirectional passing structure, iteratively computing $p\left( {{{\bf{x}}_k},{\bf{c}}_k^a\left| {{{\bf{z}}_{1:k}},{\cal D}} \right.} \right)$ and $p\left( {{{\bf{x}}_k},{\bf{c}}_k^b\left| {{{\bf{z}}_{1:K}},{\cal D}} \right.} \right)$ in the forward and backward processes, respectively. There are four explainable internal gated units in JSMT-BF and JSMT-BS: forward memory update gate (FMUG), forward filtering gate (FFG), backward memory update gate (BMUG), and backward smoothing gate (BSG). 
During filtering, FMUG updates the forward memory using the previous state and forward memory, followed by FFG compensating for the forward trend. During smoothing, BMUG updates the backward memory using the subsequent state and backward memory, followed by BSG compensating for the global trend. In addition, $p\left( {{\bf \Delta} _{k + 1}^a\left| {{\bf{c}}_{k + 1}^a,{\cal D}} \right.} \right)$ and $p\left( {{\bf{c}}_{k + 1}^a\left| {{\bf{c}}_k^a,{{\bf{x}}_k},{\cal D}} \right.} \right)$ involved in the backward smoothing stage have been computed in the forward filtering stage, so they are stored during filtering and used directly in smoothing without performing the integration operation again. Finally, the predicted, posterior and smoothed PDFs of the state are obtained by the following operations: 
\begin{align}
p\left( {{{\bf{x}}_k}\left| {{{\bf{z}}_{1:k - 1}},{\cal D}} \right.} \right) &= \int {p\left( {{{\bf{x}}_k},{\bf{c}}_k^a\left| {{{\bf{z}}_{1:k - 1}},{\cal D}} \right.} \right)d{\bf{c}}_k^a}
\label{pred_PDF}\\
p\left( {{{\bf{x}}_k}\left| {{{\bf{z}}_{1:k}},{\cal D}} \right.} \right) &= \int {p\left( {{{\bf{x}}_k},{\bf{c}}_k^a\left| {{{\bf{z}}_{1:k}},{\cal D}} \right.} \right)d{\bf{c}}_k^a}
\label{post_PDF}\\
p\left( {{{\bf{x}}_k}\left| {{{\bf{z}}_{1:K}},{\cal D}} \right.} \right) &= \int {p\left( {{{\bf{x}}_k},{\bf{c}}_k^b\left| {{{\bf{z}}_{1:K}},{\cal D}} \right.} \right)d{\bf{c}}_k^b}
\label{smoo_PDF}
\end{align}

EBRNS integrates prior model knowledge and measurements in its recursive process and uses offline data to learn unknown PDFs, including $p\left( {{\bf{c}}_k^a\left| {{{\bf{x}}_{k - 1}},{\bf{c}}_{k - 1}^a,{\cal D}} \right.} \right)$, $p\left( {{\bf{c}}_k^b\left| {{{\bf{x}}_{k + 1}},{\bf{c}}_{k + 1}^b,{\cal D}} \right.} \right)$, $p( {{{\bf \Delta} _k^a}\left| {{{\mathbf{c}}_k^a},\mathcal{D}} \right.} )$ and $p( {{{\bf \Delta} _k^b}\left| {{{\mathbf{c}}_k^b},\mathcal{D}} \right.} )$. Designing NN modules to enable offline learning is essential. Given that the computational procedure can be implemented with these distributions known, we first derive the computational implementation of EBRNS, followed by the design of the corresponding network module in the next subsection.

JSMT-BF and JSMT-BS, as general BF and BS frameworks, offer multiple implementation approaches such as Monte Carlo (MC) sampling\cite{MC1} and Gaussian approximation\cite{wang2012gaussian}. Due to its analytical properties and computational efficiency, Gaussian approximation is a fast and stable method, making it widely used in various BF and BS implementations \cite{wang_2013_Gau_filter,wang_2013_Gau_smooth,wang_2015_Gau_smooth}. Implementing EBRNS with Gaussian approximation provides good explainability and an efficient computational process. Despite the Gaussian assumption, EBRNS adapts to non-Gaussian cases by approximating non-Gaussian PDFs through the adaptive tuning of key filtering parameters \cite{yan_tsp}. The corresponding Gaussian assumptions are described below.

\noindent{Assumption 1.} The distributions of the process noise and measurement noise are Gaussian, namely, 
\begin{align}
{{\bf{w}}_{k}} \sim \mathcal  N\left( {0,{{\bf{Q}}_{k}}} \right)
\\
{{\bf{v}}_{k}} \sim \mathcal  N\left( {0,{{\bf{R}}_{k}}} \right)
\end{align}

\noindent {{Assumption 2.}} The state filtering and smoothing posterior PDFs are Gaussian:
\begin{align}
{p}\left( {{{\bf{x}}_k}\left| {{{\bf{z}}_{1:k}}} \right.} {,\mathcal{D}}\right) &= \mathcal  N\left( {{{\bf{x}}_k};{\bf{\hat x}}_{k|k},{\bf{P}}_{k|k}} \right)
\label{assum_state_up}
\\
{p}( {{{\bf{x}}_{k}}\left| {{{\bf{z}}_{1:K}}} \right.} {,\mathcal{D}}) &= \mathcal  N( {{{\bf{x}}_{k}};{\bf{\hat x}}_{k|K},{\bf{P}}_{k|K}} )
\label{assum_state_smooth}
\end{align}

\noindent{{Assumption 3.}} The measurement prediction PDF is Gaussian:
\begin{align}
p\left( {{{\bf{z}}_k}\left| {{{\bf{z}}_{1:k-1}}} \right.} {,\mathcal{D}}\right) = \mathcal  N( {{{\bf{z}}_k};{{{\bf{\hat z}}}_{k|k - 1}},{\bf{P}}_{k|k - 1}^{{z}}} )
\label{assum_meas}
\end{align}

\noindent {{Assumption 4.}} ${\bf c}_k^a$ and ${\bf c}_k^b$ obey the Gaussian distributions as:
\begin{align}
p( {{\bf c}_k^a\left| {{{\mathbf{x}}_{k - 1}},{{\bf{c}}_{k-1}^a},\mathcal{D}} \right.}) = \mathcal  N( {{\bf c}_k^a;{\bf \hat c}_k^a,{\bf{\Sigma }}_k^{ac} })
\label{assum_mem_f}
\\
p( {{\bf c}_k^b\left| {{{\mathbf{x}}_{k + 1}},{{\bf{c}}_{k+1}^b},\mathcal{D}} \right.}) = \mathcal  N( {{\bf c}_k^b;{\bf \hat c}_k^b,{\bf{\Sigma }}_k^{bc} } )
\label{assum_mem_b}
\end{align}

\noindent {{Assumption 5.}} ${\bf \Delta} _k^a$ and ${\bf \Delta} _k^a$ obey the Gaussian distributions as:
\begin{align}
&p( {{{\bf \Delta} _k^a}\left| {{{\mathbf{c}}_k^a},\mathcal{D}} \right.} ) = \mathcal  N( {{{\bf \Delta} _k^a};\hat {\bf \Delta} _k^a,{\bf{\Sigma }}_k^a } )
\label{assum_f}
\\
&p( {{{\bf \Delta} _k^b}\left| {{{\mathbf{c}}_k^b},\mathcal{D}} \right.} ) = \mathcal  N( {{{\bf \Delta} _k^b};\hat {\bf \Delta} _k^b,{\bf{\Sigma }}_k^b } )
\label{assum_b}
\end{align}

\noindent\emph{Remark 2: The general Gaussian smoother usually assumes that the distributions of ${{\bf{w}}_{k}}$, ${{\bf{v}}_{k}}$, ${{\bf{x}}_{k}}$ and ${{\bf{z}}_{k}}$ are Gaussian \cite{wang_2013_Gau_smooth}, corresponding to Assumptions 1, 2, and 3. }

In Theorems \ref{TH_GAF} and \ref{TH2}, we derive the Gaussian approximation implementations of JSMT-BF and JSMT-BS, respectively.

\begin{theorem}
\label{TH_GAF}{(Gaussian approximation implementation of the JSMT-BF)} Given Assumptions 1–5, the prediction of the state and covariance are
\begin{align}
{{{\mathbf{\hat x}}}_{k|k - 1}}=& \int {{{{f}_k}\left( {{{\mathbf{x}}_{k - 1}}} \right)}P_{k-1}^{x+}d{{\mathbf{x}}_{k - 1}}}+ \hat {\bf \Delta} _k^a
\label{Gau_state_pred}
\\
{{\mathbf{P}}_{k|k - 1}} =& \int {{{f}_k}\left( {{{\mathbf{x}}_{k - 1}}} \right)(\cdot)^{\top}P_{k-1}^{x+}d{{\mathbf{x}}_{k - 1}}}  
+ {\mathbf{P}}_k^a \notag \\-& {{{\mathbf{\hat x}}}_{k|k - 1}}{\mathbf{\hat x}}_{k|k - 1}^{\top} + {{\mathbf{Q}}_k}
\label{Gau_cov_pred}
\end{align}
and the update of the state and covariance are
\begin{align}
&{{{\mathbf{\hat x}}}_{k|k}} = {{{\mathbf{\hat x}}}_{k|k - 1}} + {\mathbf{P}}_{k|k - 1}^{xz}{\left( {{\mathbf{P}}_{k|k - 1}^z} \right)^{ - 1}}\left( {{{\mathbf{z}}_k} - {{{\mathbf{\hat z}}}_{k|k - 1}}} \right)
\label{gau_state_up}\\
&{{\mathbf{P}}_{k|k}} = {{\mathbf{P}}_{k|k - 1}} - {\mathbf{P}}_{k|k - 1}^{xz}{\left( {{\mathbf{P}}_{k|k - 1}^z} \right)^{ - 1}}{\left( {{\mathbf{P}}_{k|k - 1}^{xz}} \right)^{\top}}
\label{gau_cov_up}
\end{align}
with
\begin{align}
&{{{\mathbf{\hat z}}}_{k|k - 1}}= \int {{{h_k}\left( {{{\mathbf{x}}_k}} \right) }} P_{k}^{x-}d{{\mathbf{x}}_k}
\label{Gau_meas_pred}
\\
&{\mathbf{P}}_{k|k - 1}^z\!\!=\!\!\int \!{{h_k}\left( {{{\mathbf{x}}_k}} \right)(\cdot)^{\top}P_{k}^{x-}d{{\mathbf{x}}_k}} - {{{\mathbf{\hat z}}}_{k|k - 1}}{\mathbf{\hat z}}_{k|k - 1}^{\top}  + {{\mathbf{R}}_k}
\label{Gau_meas_cov}
\\
&{\mathbf{P}}_{k|k - 1}^{xz}={\int {{{\mathbf{x}}_k}{({{h_k}}\left( {{{\mathbf{x}}_k}} \right))^{\top}}}}P_{k}^{x-}d{{\mathbf{x}}_k} - {{{\mathbf{\hat x}}}_{k|k - 1}}{\mathbf{\hat z}}_{k|k - 1}^{\top}
\label{Gau_xz_cov}
\\
&P_{k-1}^{x+}=N( {{{\mathbf{x}}_{k - 1}};{{{\mathbf{\hat x}}}_{k - 1|k - 1}},{{\mathbf{P}}_{k - 1|k - 1}}} )\notag
\\
&P_{k}^{x-}=N( {{{\mathbf{x}}_{k}};{{{\mathbf{\hat x}}}_{k|k - 1}},{{\mathbf{P}}_{k|k - 1}}} )
\notag
\end{align}
\end{theorem}
\begin{proof}
See the proof of Theorem 2 in our previous work \cite{yan_tsp}, where the compensation for the observation model is not considered.
\end{proof} 

\begin{theorem}\label{TH2}{(Gaussian approximation implementation of the JSMT-BS)} Under Assumptions 1–5, given $P_k^{x + }=N\left( {{{\bf{x}}_{k|k}};{{{\bf{\hat x}}}_{k|k}},{{\bf{P}}_{k|k}}} \right)$, the state and covariance smoothing are
\begin{align}
&{{{\bf{\hat x}}}_{k|K}} = {{{\bf{\hat x}}}_{k|k}} + {{\bf{G}}_k}\left( {{{{\bf{\hat x}}}_{k + 1|K}} - {\bf{\hat x}}_{k + 1|k}^b} \right)
\label{smooth_x}\\
&{{\bf{P}}_{k|K}} = {{\bf{P}}_{k|k}} + {{\bf{G}}_k}\left( {{{\bf{P}}_{k + 1|K}} - {\bf{P}}_{k + 1|k}^b} \right){\bf{G}}_k^{\rm{T}}
\label{smooth_P}
\end{align}
with
\begin{align}
&{{{\bf{\hat x}}}_{k + 1|k}}^b = {{{\mathbf{\hat x}}}_{k|k - 1}} + \hat {\bf \Delta} _{k+1}^b
\label{Back_state_pred}
\\
&{{\bf{P}}_{k + 1|k}^b}= {{\mathbf{P}}_{k|k - 1}} + {\bf{\Sigma }}_{k+1}^b
\label{Back_cov_pred}
\\
&{{\bf{G}}_k} = {{\bf{C}}_{k,k + 1}}{\left( {{\bf{P}}_{k + 1|k}^b} \right)^{ - 1}}
\label{Back_gain}
\\
&{{\bf{C}}_{k,k + 1}} = \int\!\! {{{\bf{x}}_k}{{\left( {f_{k+1}\left( {{{\bf{x}}_k}} \right)} \right)}^{\rm{T}}} } P_k^{x + }d{{\bf{x}}_k} + {{\bf{\hat x}}_{k|k}}{\left( {f_{k+1}\left( {{{{\bf{\hat x}}}_{k|k}}} \right)} \right)^{\rm{T}}}
\label{Back_C}
\end{align}
\end{theorem}
\begin{proof}
See Appendix \ref{Proof_B}.
\end{proof} 
\begin{figure*}[t]
\centering
\includegraphics[width=1.\linewidth]{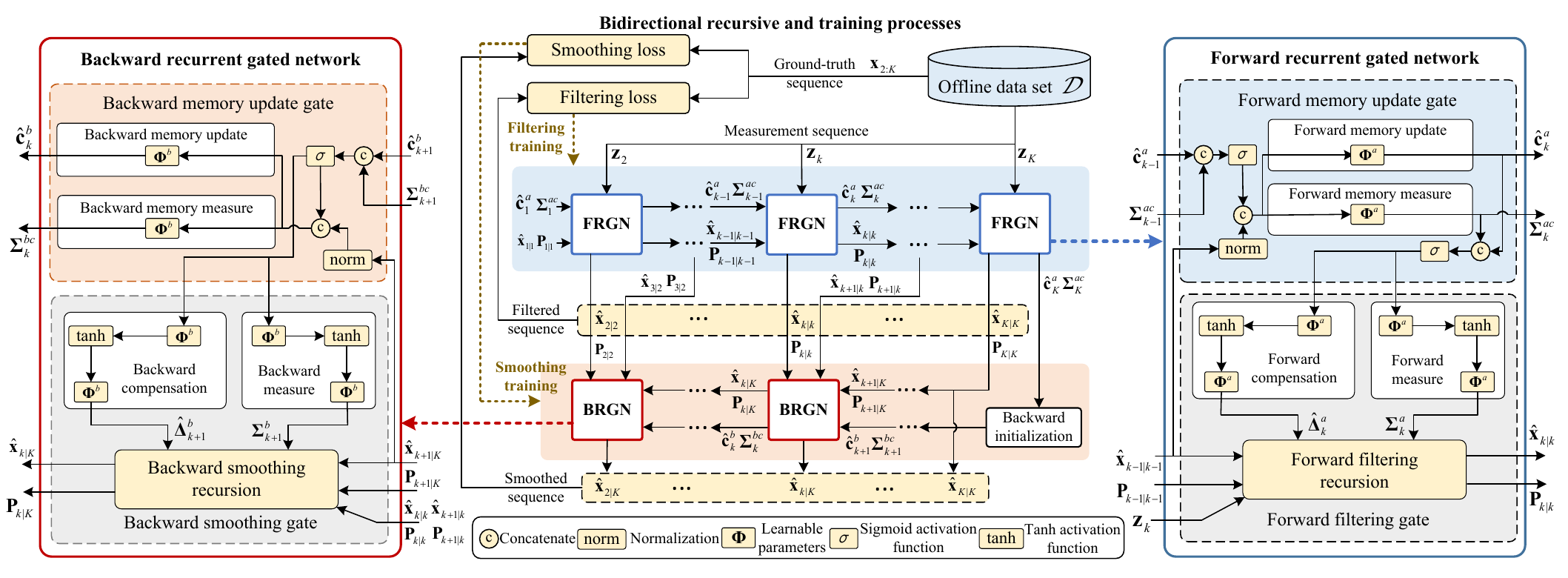}
\caption{{The overall architecture and two-stage offline training process of EBRNS.}}
\label{framework}
\end{figure*}

To compute the nonlinear integrals in Theorems \ref{TH_GAF} and \ref{TH2}, several fast Gaussian approximation methods are available \cite{wang2012gaussian}. In this paper, we exemplify the computation of these integrals with the first-order Taylor expansion. We first denote the Jacobian matrices of ${f}_k$ and ${h}_k$ as ${\bf F}_k$ and ${\bf H}_k$, respectively. Following the function approximation approach of the extended Kalman filter, Eqs. (\ref{Gau_state_pred}), (\ref{Gau_cov_pred}), (\ref{Gau_meas_pred})-(\ref{Gau_xz_cov}), and (\ref{Back_C}) are replaced with:
\begin{align}
{{{\mathbf{\hat x}}}_{k|k - 1}} =& {{f}_k}\left( {{{{\mathbf{\hat x}}}_{k - 1|k - 1}}} \right) + \hat {\bf \Delta} _k^a
\label{ekf_state_pred}\\
{{\mathbf{P}}_{k|k - 1}} =& {{{\mathbf{F}}}_k}{{\mathbf{P}}_{k - 1|k - 1}}{\mathbf{F}}_k^{\top} + {{\mathbf{Q}}_k} + {\mathbf{P}}_k^a
\label{ekf_cov_pred}\\
{{{\mathbf{\hat z}}}_{k|k - 1}} =& {{h}_k}\left( {{{{\mathbf{\hat x}}}_{k|k - 1}}} \right)
\label{ekf_meas_pred}\\
{\mathbf{P}}_{k|k - 1}^z =& {{\mathbf{H}}_k}{{\mathbf{P}}_{k|k - 1}}{\mathbf{H}}_k^{\top}{\text{ + }}{{\mathbf{R}}_k}
\label{ekf_measP_pred}\\
{\mathbf{P}}_{k|k - 1}^{xz} =& {{\mathbf{P}}_{k|k - 1}}{\mathbf{H}}_k^{\top}
\label{ekf_xzP}\\
{{\bf{C}}_{k,k + 1}}=& {\bf P}_{k|k}{\bf F}_{k+1}^\top
\end{align}

Through the Gaussian approximation, we convert the learning requirement of EBRNS for unknown PDFs into the estimation of their first- and second-order moments. To achieve this, we need to construct NN modules to estimate these moments for the gated units in EBRNS via offline learning.

\subsection{Internal network design}
Fig. \ref{framework} shows the overall architecture of EBRNS and the internal structure of its gated networks. The forward recurrent gated network (FRGN) is derived from the JSMT-BF, which iteratively computes the forward memory and implements filtering. Then, the backward recurrent gated network, which is derived from the JSMT-BS, is used to iteratively compute the backward memory and implement smoothing.
Based on BE theory, EBRNS introduces the prior model information in an explainable computational process and learns its NN-based internal gated units using offline data to compensate for the evolutionary trend. The good explainability eliminates the need for complex network structures, and we construct these network modules as two-layer NNs with a hyperbolic tangent activation function, i.e.,
\begin{align}
{\cal F}\left( {{\bf i},j} \right) = {\omega ^{j2}}\left( {\tanh \left( {{\omega ^{j1}}{\bf i} + {{\mathbf{b}}^{j1}}} \right)} \right) + {{\mathbf{b}}^{j2}}
\label{NN_structure}
\end{align}
where $\rm tanh$ is the hyperbolic tangent activation function, $j$ is the index, ${\bf i}$ is the input vector, $\omega$ and $\bf b$ are the weights and biases of the NN, respectively. 

The inputs to each NN module are described as follows, with relevant notations explained: $\text{flat}$ represents the flattening of a matrix to a vector, $\sigma$ is the sigmoid activation function, ${\rm concat}$ denotes the concatenation operation, ${\rm norm}$ indicates the maximal normalization.

\subsubsection{Input of the NN modules in FMUG} The NN modules are used to estimate ${\bf \hat c}_k^a$ and ${\bf{\Sigma }}_k^{ac}$ in FMUG. According to $p( {{\bf c}_k^a\left| {{{\mathbf{x}}_{k - 1}},{{\mathbf{c}}_{k-1}^a},\mathcal{D}} \right.} ) $, the corresponding input vector is ${\bf i}_k^{ac} = {\text{concat}}[ {\sigma ( {{\text{concat}}[ {{{{\mathbf{\hat c}}}_{k - 1}^a},{\text{flat}}( {{\mathbf{\Sigma }}_{k - 1}^{ac}} )} ]} ),{\rm norm}( {{{{\mathbf{\hat x}}}_{k - 1|k - 1}}} )} ]$.

\subsubsection{Input of the NN modules in FFG} 
The NN modules are used to estimate $\hat {\bf \Delta} _k^a$ and ${\mathbf{\sigma}}_{k}^a$ in FFG. According to $p( {{{\bf \Delta} _k^a}\left| {{{\mathbf{c}}_k^a},\mathcal{D}} \right.})$, the corresponding input vector is ${\mathbf{i}}_k^a = {\sigma ( {{\text{concat}}[ {{{{\mathbf{\hat c}}}_k^a},{\text{flat}}( {{\mathbf{\Sigma}}_k^{ac}} )} ]} )}$.

\subsubsection{Input of the NN modules in BMUG} The NN modules are used to estimate ${\bf \hat c}_k^b$ and ${\bf{\Sigma }}_k^{bc}$ in BMUG. According to $p( {{\bf c}_k^b\left| {{{\mathbf{x}}_{k + 1}},{{\mathbf{c}}_{k+1}^b},\mathcal{D}} \right.} ) $, the corresponding input vector is ${\bf i}_k^{bc} = {\text{concat}}[ {\sigma ( {{\text{concat}}[ {{{{\mathbf{\hat c}}}_{k + 1}^b},{\text{flat}}( {{\mathbf{\Sigma }}_{k + 1}^{bc}} )} ]} ),{\rm norm}( {{{{\mathbf{\hat x}}}_{k + 1|K}}} )} ]$.

\subsubsection{Input of the NN modules in BSG} 
The NN modules are used to estimate $\hat {\bf \Delta} _k^b$ and ${\mathbf{\Sigma}}_{k}^b$ in BSG. According to $p( {{{\bf \Delta} _k^b}\left| {{{\mathbf{c}}_k^b},\mathcal{D}} \right.})$, the corresponding input vector is ${\mathbf{i}}_k^b = {\sigma ( {{\text{concat}}[ {{{{\mathbf{\hat c}}}_k^b},{\text{flat}}( {{\mathbf{\Sigma}}_k^{bc}} )} ]} )}$.

\begin{table}[htp]
\renewcommand{\arraystretch}{1.5}
\centering
\caption{Mappings formed by the NN modules in each gated unit.}
\setlength{\tabcolsep}{3.mm}
\label{NNmapping}
\begin{tabular}{ccccccc}
\hline
                {\textbf{Gated unit}}  & {\textbf{NN module}}   & {\textbf{Mapping}} \\ \hline
\multirow{2}{*}{FMUG} &Forward memory update  &${{{\mathbf{\hat c}}}_k^a} = F( {{\mathbf{i}}_k^{ac},ac1} )$  \\
                  &Forward memory measure& ${\mathbf{\Sigma}}_k^{ac} = F( {{\mathbf{i}}_k^{ac},ac2} )$ \\ \hline 
\multirow{2}{*}{FFG} &Forward compensation &$\hat \Delta _k^a = F( {{\mathbf{i}}_k^a,a1} )$ \\
                  &Forward measure  &${\mathbf{\Sigma}}_{k}^{a}=F( {{\mathbf{i}}_k^a,a2} )$  \\ \hline 
\multirow{2}{*}{BMUG} &Backward memory update  &${{{\mathbf{\hat c}}}_k^b} = F( {{\mathbf{i}}_k^{bc},bc1} )$  \\
                  &Backward memory measure& ${\mathbf{\Sigma}}_k^{bc} = F( {{\mathbf{i}}_k^{bc},bc2} )$ \\ \hline 
\multirow{2}{*}{BSG} & Backward compensation & $\hat \Delta _k^b =F( {{\mathbf{i}}_k^b,b1} )$ \\
                  &Backward measure& ${\mathbf{\Sigma}}_{k}^b=F( {{\mathbf{i}}_k^b,b2} )$ \\ \hline 
\end{tabular}
\end{table}

The corresponding mappings of the NN modules in EBRNS are shown in Tab. \ref{NNmapping}. Supported by good physical explainability, the gated units have their own well-defined functions, and the evolutionary trends are compensated by them at different stages of filtering and smoothing. In this paper, we further explore the natural integration of BE theory and DL in state smoothing. The EBRNS, which contains two gated networks, can also be considered a Bi-RNN. Different from the widely used general Bi-RNN, EBRNS is specifically designed for state smoothing of the dynamic system, and its bidirectional passing structure and the structure of its internal gated networks are derived from BE theory, which has good physical explainability and theoretical guarantees. It introduces a novel perspective for model-based DL, i.e., exploring the explainable underlying network structure along with the explainable computational process.

\subsection{Algorithm training}
As shown in Fig. \ref{framework}, EBRNS undergoes supervised training in two stages to extract forward and backward evolutionary correlation information from ${\cal D}$. In the first stage, learnable parameters in BRGN are fixed, while FRGN is trained end-to-end based on filtering results. In the second stage, learnable parameters in FRGN are fixed, while BRGN is trained end-to-end based on smoothing results. This staged learning aligns with the optimization order of EBRNS, ensuring stable algorithm training. Although NN modules in FRGN and BRGN compute intermediate parameters rather than directly estimating states, they are trained by using loss functions based on the estimated states and corresponding ground-truth. The inputs, outputs, and placement of sub-network modules in the gated structure of FRGN and BRGN have explicit physical meanings. Through end-to-end training, each network module learns the optimal parameters required for specific functions based on their placement and role in the overall network and their impact on the final state estimation.

Based on the minimum mean square error criterion, two mean square error (MSE) loss functions with L2 regularization are constructed for the two training stages:
\begin{flalign}
{l_{i}^a }\left ( \Phi^a \right ) =\frac{1}{K} \sum_{k=2}^{K} \left \| {{\bf \hat x}_{k|k}^{i}\left( \Phi^a \right) -{\bf x}_{k}^{i}} \right \|^{2}+\tau^a \left \| \Phi^a \right \| ^{2}  
\label{loss_i_a}
\\
{l_{i}^b }\left ( \Phi^b \right ) =\frac{1}{K} \sum_{k=2}^{K} \left \| {{\bf \hat x}_{k|K}^{i}\left( \Phi^b \right) -{\bf x}_{k}^{i}} \right \|^{2}+\tau^b \left \| \Phi^b \right \| ^{2}  
\label{loss_i_b}
\end{flalign}
where ${l_{i}^a }$ and ${l_{i}^a }$ are the filtering and smoothing losses, respectively; $\tau^a$ and $\tau^b$ are the regularization weights; $i \in \{1,2, \cdots ,I\}$ denotes the sample index; $\Phi^a$ and $\Phi^b$ are the learnable parameter sets of FRGN and BRGN, respectively. 

The end-to-end training of the NN modules in EBRNS is performed using stochastic gradient descent. Since all operations in the filtering and smoothing process implemented by Gaussian approximation are differentiable, these gradients can be computed. It is very difficult to compute these gradients manually given the complex nonlinear operations that may be involved, so we use the automatic differentiation tool \cite{tensorflow} to compute these gradients. 
Specifically, we employ a mini-batch stochastic gradient descent algorithm, i.e., for each batch indexed by $n$, we select $J < I$ trajectories denoted as $i_{1}^{n}, i_{2}^{n}, \ldots, i_{J}^{n}$, respectively. The corresponding mini-batch losses of Eqs. (\ref{loss_i_a}) and (\ref{loss_i_b}) are:  
\begin{flalign}
L_n^a\left (\Phi^a \right ) &=\frac{1}{J} \sum_{j=1}^{J} {l_{i_{j}^{n}}^a\left ( \Phi^a \right )  } 
\\
L_n^b\left (\Phi^b \right ) &=\frac{1}{J} \sum_{j=1}^{J} {l_{i_{j}^{n}}^b\left ( \Phi^b \right )  } 
\label{loss_bt}
\end{flalign}
Since the gated networks in the forward and backward passes are trained separately, this allows the EBRNS to be trained using the same algorithm as a general RNN. Here, the EBRNS is trained using the backpropagation through  time algorithm \cite{bptt}, an efficient algorithm for temporal DL model training.

\section{Experiments}
In this section, the smoothing performance of EBRNS is evaluated in two representative state estimation tasks under the uncertain model. Utilizing real-world datasets with differing state dimensions, we compare EBRNS with various benchmarks and state-of-the-art smoothing methods. Our evaluation focuses on estimation accuracy, data dependency, and parameter utilization efficiency. We employ two metrics to evaluate estimation accuracy: the root MSE (RMSE) and mean RMSE, computed through $M$ times MC runs. They are specifically computed as
\begin{align}
&{{R}}_{{{x}},k} = \sqrt {{1 \over M}\sum\limits_{i = 1}^{M} {\left( {{{\bf{x}}_{k}^{(i)}} - {{{\bf{\hat x}}}_{k|K}^{(i)}}} \right){{\left( {{{\bf{x}}_{i,k}} - {{{\bf{\hat x}}}_{k|K}^{(i)}}} \right)}^\top}} }
\label{RMSE_1}\\
&{{R}}^{\rm mean}_x = \sqrt {{1 \over {KM}}\sum\limits_{i = 1}^{M} {\sum\limits_{k = 1}^K {\left( {{{\bf{x}}_{k}^{(i)}} - {{{\bf{\hat x}}}_{k|K}^{(i)}}} \right){{\left( {{{\bf{x}}_{k}^{(i)}} - {{{\bf{\hat x}}}_{k|K}^{(i)}}} \right)}^\top}} } }
\label{RMSE_2}
\end{align}

\subsection{Temperature series smoothing}
\label{time_sreies_ex}
Estimating temperature from noisy observations is crucial for weather prediction and climate research. This experiment assesses EBRNS's performance in smoothing real-world temperature series characterized by significant periodicity, trend, and autocorrelation properties. In addition, we demonstrate the low dependency of EBRNS on the training data.

\textbf{Experimental setups.} 
The experiment employs a weather dataset containing detailed hourly and daily summaries of weather conditions in the Szeged area from 2006 to 2016 \cite{dataset_weather}. This dataset, which is commonly employed in various time-series analysis tasks, offers several weather variables, including temperature, humidity, and wind speed\cite{time_data_appl_1,time_data_appl_2}. It contains $96,000$ frames of temperature data with a one-hour interval between each frame. We divide it with the length of 48 hours, yielding a total of 2000 samples, with 1,400 for training, 400 for validation, and 200 for testing (in a proportion of 7:2:1). The estimated state ${x_k}$ is the temperature and the corresponding measurement is
\begin{align}
{z_k} = {x_k} + {v_k}
\label{temp_meas}
\end{align}
where the noise $v_k \sim N(0,\sigma_v^2)$. To evaluate the smoothing performance and stability of EBRNS under different noise levels, we set up four noise levels of $\sigma_v =2^\circ C$, $4^\circ C$, $6^\circ C$, and $8^\circ C$.

For setups of EBRNS, its nominal state evolution function is $f(x_k)=x_k$ and its nominal measurement model is set as Eq. (\ref{temp_meas}). The nominal state evolution function and the real temperature evolution are evidently mismatched, requiring compensation that addresses the periodicity and trend of the state evolution. Given the low dimensionality of the state, a limited number of learnable parameters are assigned to EBRNS to prevent overfitting. Specifically, both the number of hidden nodes in the NN modules and the dimension of the memory vectors are set to $32$.

The comparison methods are outlined as follows: 1) Bi-LSTM and Bi-GRU: These are two representative bidirectional gated RNNs, each adopting a three-layer stacked structure with $64$ hidden nodes in each layer. 2) Transformer \cite{Transformer}: A three-layer encoding-decoding structure with self-attention and mutual-attention mechanisms is used, which has four attention heads and $64$ hidden layer nodes. 3)Bi-Mamba: A SOTA temporal DL model, constructed based on bidirectional Mamba blocks \cite{BiMamba}. We used a single Bi-Mamba block with the number of $64$ hidden layer nodes in each Mamba block. Note that the above DL modes map the measurement directly to the state. 4) KS: A standard KS with nominal model settings identical to EBRNS. 4) EGBRNN \cite{yan_tsp}: A Bayesian gated RNN for filtering, which is equivalent to a version of EBRNS that contains only forward filtering and has the same parameter settings as EBRNS.

\begin{figure}[t]
\centering
\includegraphics[width=0.95\linewidth]{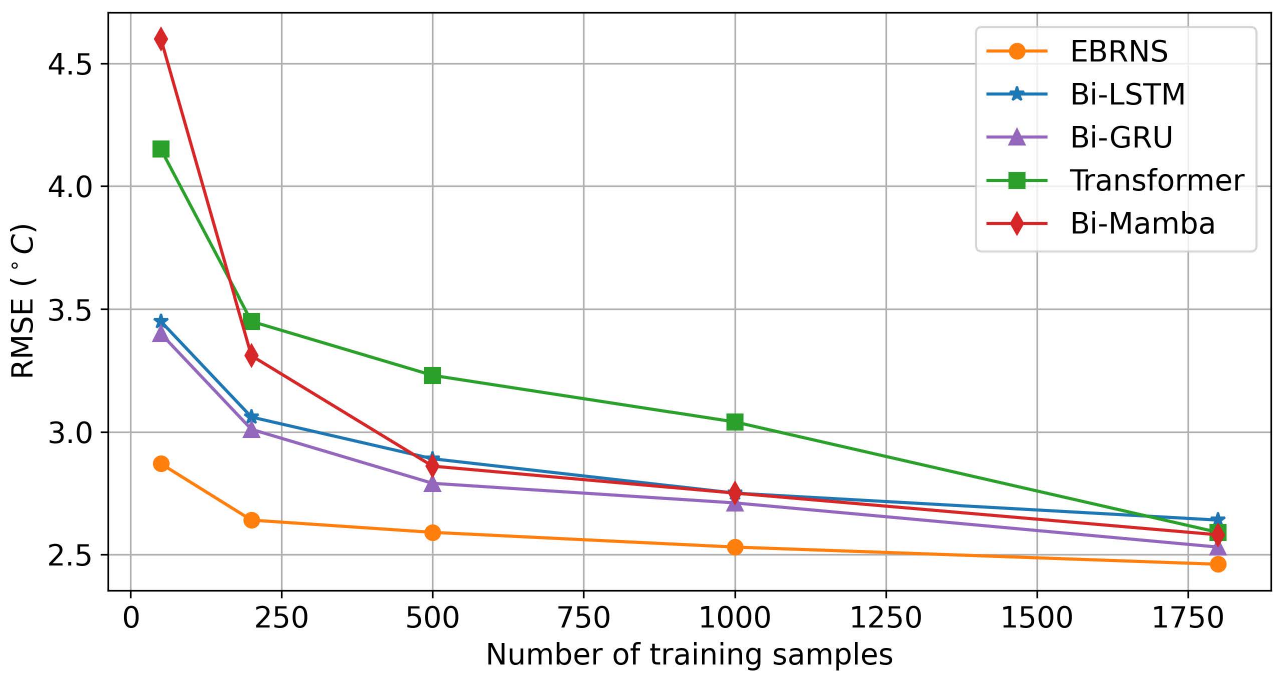}
\caption{Mean RMSE ($^\circ C$) for each method with different numbers of training samples.}
\label{train_sample_1D}
\end{figure}

\begin{table}[t]
\renewcommand{\arraystretch}{1.6}
\setlength\tabcolsep{4pt}
\begin{center}
\caption{Mean RMSE ($^\circ C$) on test set with training on 500 samples.}
\label{mean_RMSE_1D_test}
\setlength{\tabcolsep}{2.mm}
\begin{tabular}{ ccccccc }
\hline
 {\makecell{Noise level}} & {\makecell{$\sigma _v=2^\circ C$}}&{\makecell{$\sigma _v=4^\circ C$}}& {\makecell{$\sigma _v=6^\circ C$}}& {\makecell{$\sigma _v=8^\circ C$}}\\
\hline
{\makecell{{Bi-LSTM}}}&1.17&1.80&2.37&2.88\\
\hline
{\makecell{{Bi-GRU}}}&1.14&1.78&2.30&2.79\\
\hline
{\makecell{{Transformer}}}&1.59&2.14&2.71&3.23\\
\hline
{\makecell{{Bi-Mamba}}}&1.10&1.82&2.36&2.86\\
\hline
{\makecell{{KS}}}&1.12&1.78&2.31&2.80\\
\hline
{\makecell{{EGBRNN}}}&1.37&{2.11}&{2.67}&{3.07}\\
\hline
{\makecell{\textbf{EBRNS}}}&\textbf{1.06}&\textbf{1.63}&\textbf{2.06}&\textbf{2.48}\\
\hline
\end{tabular}
\end{center}
\end{table}

\begin{figure}[t]
\centering
\includegraphics[width=0.95\linewidth]{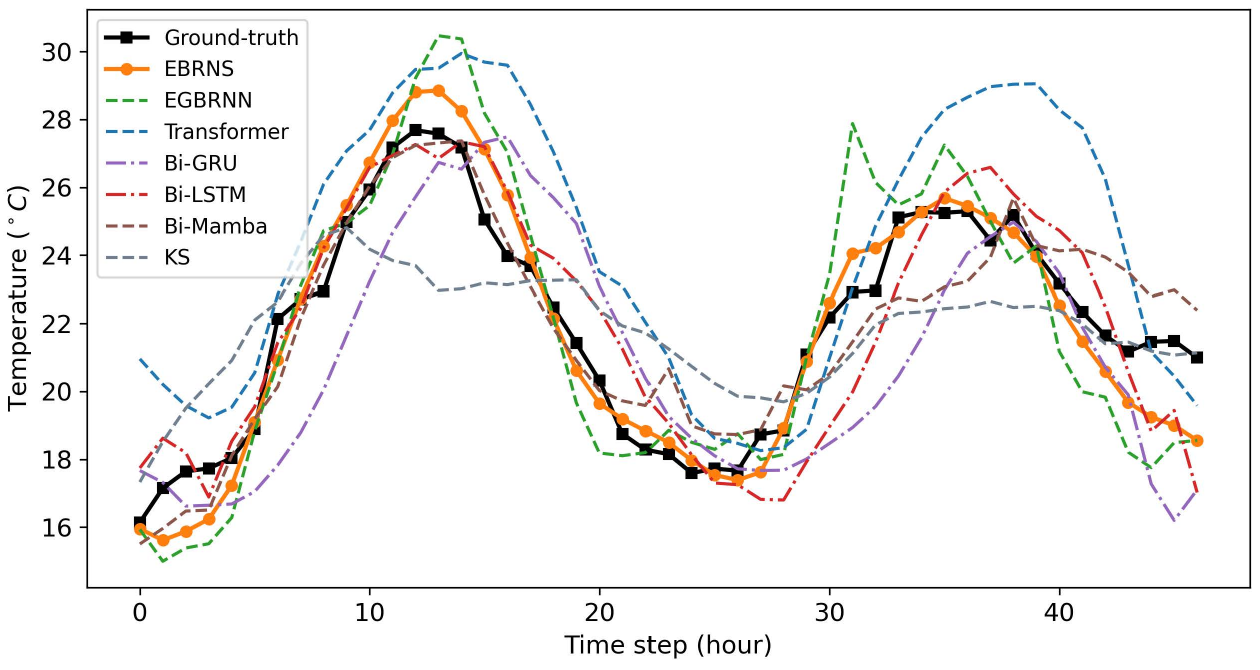}
\caption{Smoothing result for a single test temperature sequence.}
\label{Res_1D}
\end{figure}
\begin{figure}[t]
\centering
\includegraphics[width=0.95\linewidth]{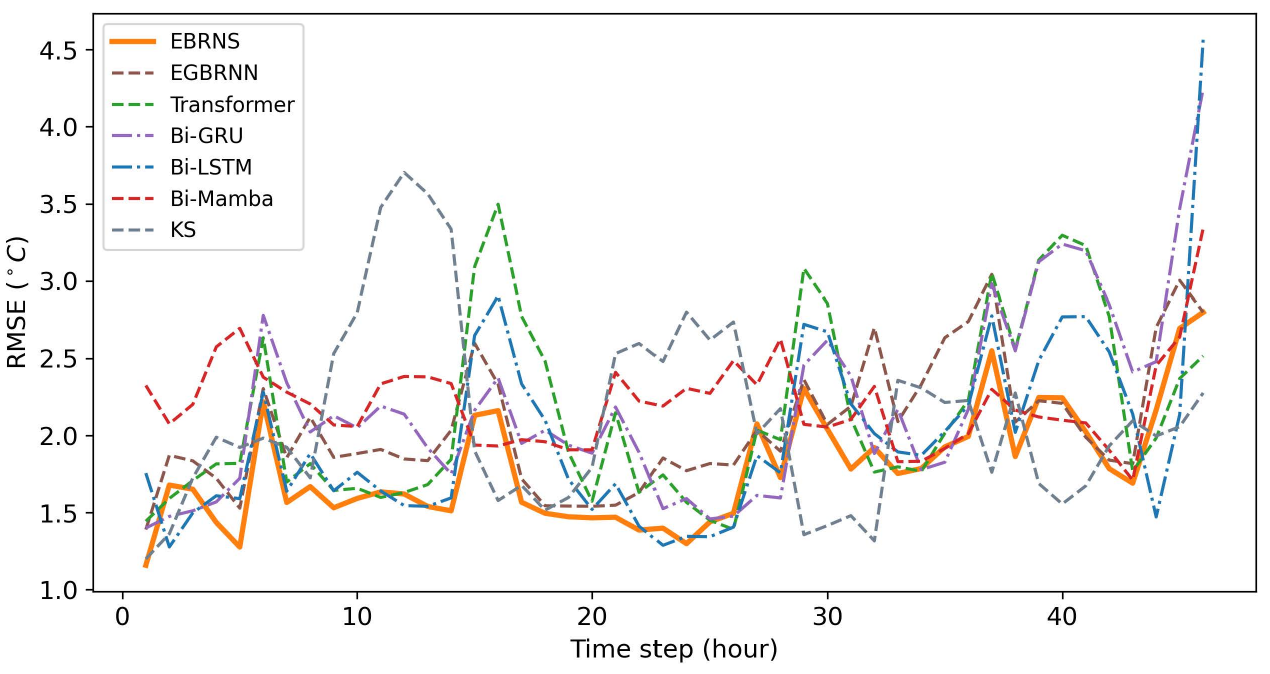}
\caption{Smoothing RMSE ($^\circ C$) of 100 MC estimation for a single test temperature sequence.}
\label{RMSE_1D}
\end{figure}

\textbf{Experimental results and analysis.} 
Considering $\sigma_v=8^\circ C$, we first evaluate the estimation performance of each method under different training data scales, and the mean smoothing RMSEs of these methods are shown in Fig. \ref{train_sample_1D}. With a sufficient number of training samples, the smoothing accuracy of EBRNS is close to that of several temporal DL models. However, with diminishing training sample sizes, conventional time-series DL models witness a significant rise in smoothing RMSE, contrasting with EBRNS's sustained high accuracy even with limited training data. This highlights that the completely data-driven DL models require substantial training data to adequately  capture the temperature evolution and noise statistics information. In contrast, EBRNS learns efficiently by its explainable computational architecture, reasonably introducing the prior noisy statistical information, making it capable of achieving highly accurate estimation even with limited samples.

Tab. \ref{mean_RMSE_1D_test} shows the mean smoothing RMSE for each method at different noise levels with a moderate training data scale (500 training samples). EBRNS provides accurate estimations across varying noise levels, particularly excelling as noise increases. In contrast, MB KS relies solely on prior model information and lacks learning ability. While DL models can leverage offline data, their lack of physical explainability results in inefficient data usage and lower smoothing accuracy. In addition, EGBRNN, which only captures forward state correlations, shows higher RMSE, illustrating the enhancements of EBRNS through global information capture.

We detailed the smoothing performance for a single test sample with $\sigma _v=6^\circ C$, with the corresponding individual smoothing results and 100 MC smoothing RMSE shown in Figs. \ref{Res_1D} and \ref{RMSE_1D}, respectively. As shown in Fig. \ref{Res_1D}, the 48-hour temperature evolution curve is periodic and trending. EBRNS achieves stable and accurate state smoothing by capturing the periodicity and trend implied in global state information through learning from offline data. As shown in Fig. \ref{RMSE_1D}, the EBRNS has both low steady-state errors and peak errors.

\subsection{Landing aircraft trajectory smoothing}
This experiment evaluates the smoothing performance of EBRNS on real-world aircraft landing trajectory data. During landing, state evolution is jointly influenced by forward information from the starting point and backward information from the destination. Under the nonlinear radar observation model, we showcase the ability of EBRNS to capture global state evolutionary correlations. We also assess its performance across different parameter scales.

\textbf{Experimental setups.} 
Following the experimental setup in \cite{yan_tsp}, we use trajectory data from the public dataset in \cite{air_data}, which includes flight records from January to March 2006 in Northern California TRACON. Aircraft trajectories heading east and landing on a specific runway are extracted and interpolated to a fixed sampling interval of ${\bf \Delta} t = 4$ by using cubic spline interpolation. The flight duration is standardized to $K = 200$, discarding shorter and truncating longer trajectories, resulting in 260 trajectories. We divided the dataset according to the usual proportion of 7:2:1, with the training set containing 183 trajectories, the validation set containing 52 trajectories, and the test set containing 26 trajectories. The targets are tracked in the two-dimensional Cartesian coordinate system. The state vector is ${{\bf{x}}_k} = [p_k^x,p_k^y,v_k^x,v_k^y]^{\top}$ with position $[p_k^x,p_k^y]$ and velocity $[v_k^x,v_k^y]$. A general nonlinear radar observation model is used for the generation of measurements. The measurement vector is ${{\bf{z}}_k} = [{\eta _k},{\alpha _k}]^{\top}$ with the ${{\eta_{k}}}$ and ${\alpha _{k}}$ are the radial distance and azimuth angle, respectively. The state-to-measurement function is
\begin{equation}
h({{\bf{x}}_{k}}){\rm{ = }}{\left[ {\sqrt {(p^x_{k})^2 + (p^y_{k})^2} ,\arctan \left( {{(p^y_{k}) \over (p^x_{k})}} \right)} \right]^{\top}}
\label{radar_h}
\end{equation}
and the observation noise is ${{\bf{v}}} = {\left[ {\rho_{\eta},\rho_\alpha } \right]^\top}$ with ${\rho_{\eta}} \sim \mathcal  N\left( {{\rho_{\eta}};0,{\sigma _{\eta}}^2} \right)$ and ${\rho_\alpha } \sim \mathcal N\left( {{\rho_\alpha };0,{\sigma _\alpha }^2} \right)$. 

The nominal observation model is accurately defined since ATC radars are typically reliable cooperative sensors. The measurement noise covariance is ${\mathbf{R}} = {\rm{diag}}\left( {\sigma _{\eta}^2,\sigma _\alpha ^2} \right)$. Given the challenges in accurately modeling aircraft motion, the nominal dynamic model is a constant velocity model, and the corresponding nominal state evolution function is
\begin{align}
    {f_k\left({\bf{x}}_{k-1}\right)}&=\begin{bmatrix}
   1 & 0 & {{\Delta} t} & 0  \cr
   0 & 1 & 0 & {{\Delta} t}  \cr
   0 & 0 & 1 & 0  \cr
   0 & 0 & 0 & 1  \cr
\end{bmatrix}{\bf{x}}_{k-1}
\label{CV_model}
\end{align}
where the sampling interval ${\Delta} t = 4$ and the nominal process noise covariance is ${\bf Q} = {\rm{diag}}\left( {{q^2},{q^2},{q^2},{q^2}} \right)$ with $q^2=10$.

The setups for the comparison methods are as follows: 1) Transformer and Bi-Mamba: Measurements in the polar coordinate system are converted to Cartesian coordinates and then input to the DL model for smoothing. Both Transformer and Bi-Mamba models have the same structure as in \ref{time_sreies_ex}, and the number of hidden layer nodes is set to $64$. 2) EKS: A standard EKS for nonlinear state smoothing with the same nominal model settings as EBRNS. 3) EGBRNN: Uses the same settings as EBRNS. 4) RTSNet: A SOTA data-assisted state smoothing algorithm that synthesizes prior model knowledge and offline data, and its setting references \cite{RTSNet}.

\textbf{Experimental results and analysis.}
Tabs. \ref{Pos_RMSE_tab} and \ref{Vel_RMSE_tab} show the mean position and velocity RMSEs for EBRNS and the comparison methods at varying noise levels, respectively. Pure DL models struggle to capture state evolution with limited data due to the complexity and dynamics of aircraft motion, resulting in high RMSEs. High-velocity RMSEs for these methods indicate poor model learning capability. While EKS achieves better smoothing accuracy by leveraging prior model information, it lacks offline learning. RTSNet, integrating model and data advantages, achieves precise position estimates but is not very good at velocity estimation due to its gain-learning framework, which does not directly compensate for state evolution trends. EGBRNN, focusing only on forward state correlation and ignoring backward information, is less effective. In contrast, EBRNS effectively integrates prior model information and offline data within a robust architecture, capturing both forward and backward state evolution correlations, leading to accurate position and velocity estimation.

\begin{figure}[t]
\centering
\includegraphics[width=0.95\linewidth]{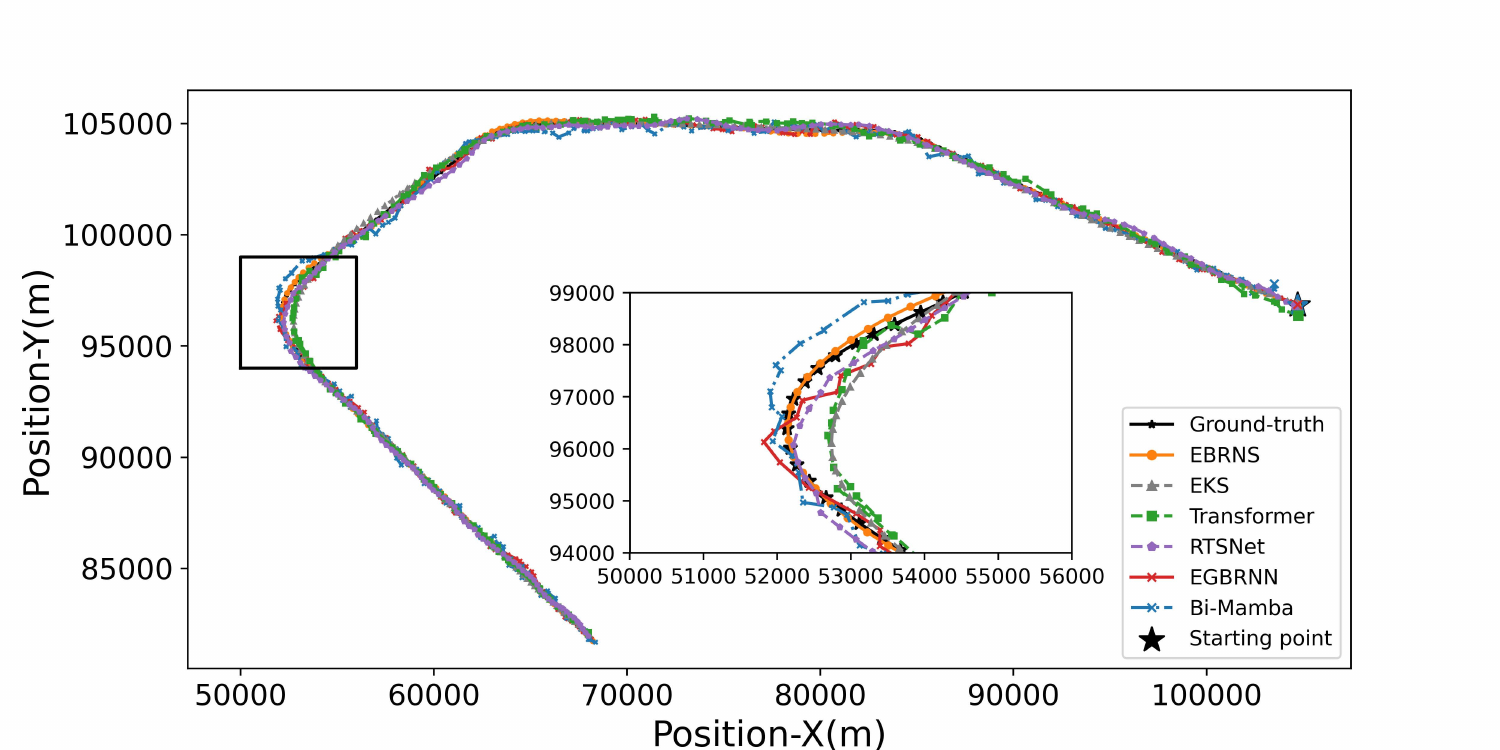}
\caption{Tracked trajectories of each method for a single test sample.}
\label{single_track}
\end{figure}

\begin{figure}[t]
\centering
\includegraphics[width=0.95\linewidth]{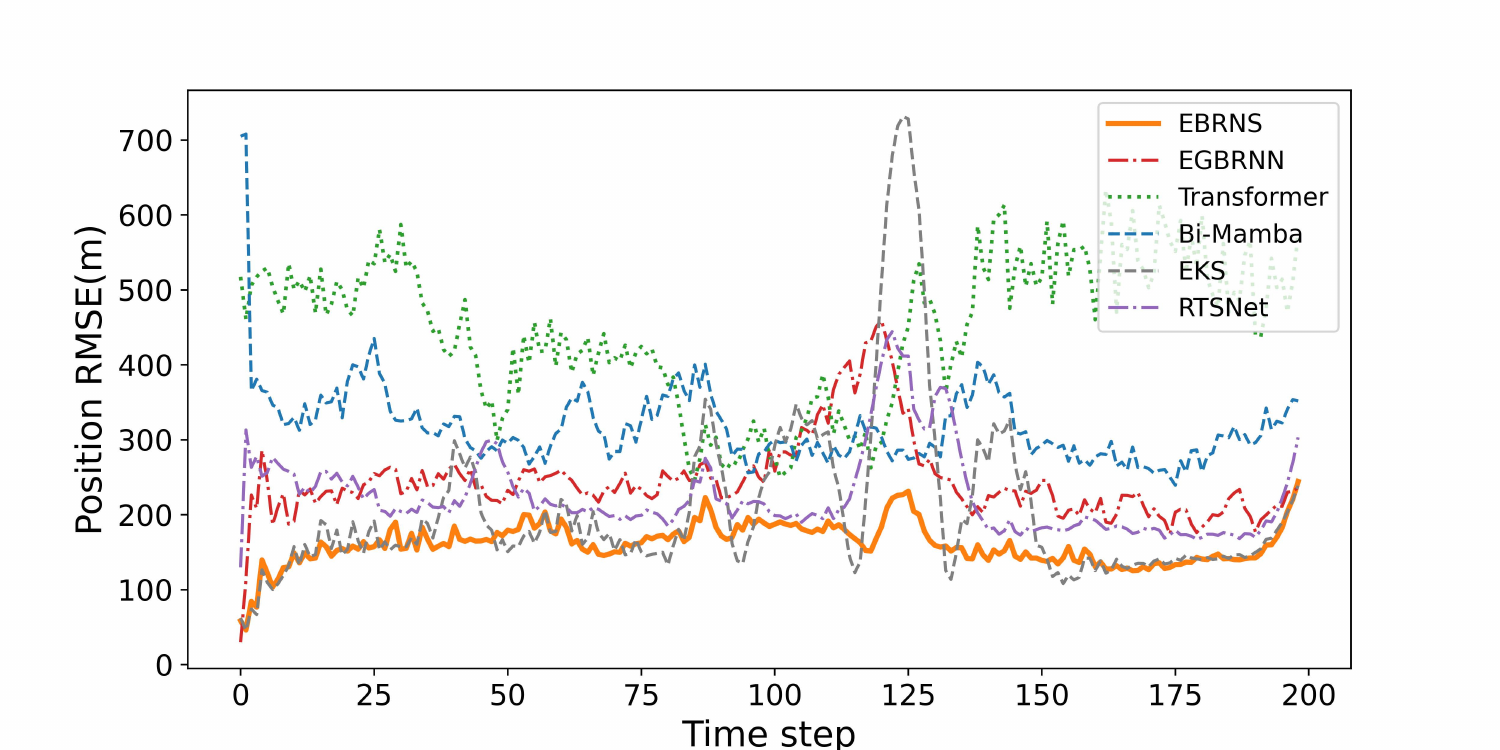}
\caption{Position RMSE (m) of 100 MC tracking for a single test sample.}
\label{RMSE_pos}
\end{figure}

Figs. \ref{single_track} and \ref{RMSE_pos} show the smoothing results and 100 MC simulation RMSEs for each method at a noise level of $(0.3^\circ,150\text{m})$, respectively. Based on the introduction of the prior model knowledge, EBRNS captures forward evolutionary information from the starting point and backward information from the destination, producing a trajectory that aligns well with the actual one and achieves low peak and steady-state errors. In contrast, EGBRNN captures only forward information, and RTSNet fails to directly compensate for evolutionary trends or introduce prior noise statistics, resulting in higher errors. EKS, lacking model learning capability, shows high peak errors when the prior model is mismatched, and pure DL methods exhibit high overall errors.
\begin{table}[t]
\renewcommand{\arraystretch}{1.8}
\setlength\tabcolsep{4pt}
\begin{center}
\caption{Mean position RMSE (m) on the test set.}
\label{Pos_RMSE_tab}
\setlength{\tabcolsep}{.8mm}
\begin{tabular}{ ccccccc }
\hline
 {\makecell{Noise level\\($\sigma_\alpha,\sigma_{\eta}$)}} & {\makecell{$ (0.1^\circ,50\text{m}) $}}&{\makecell{$ (0.15^\circ,100\text{m}) $}}& {\makecell{$ (0.2^\circ,100\text{m}) $}}& {\makecell{$ (0.3^\circ,150\text{m}) $}}\\
\hline
{\makecell{{EKS}}}&78.7&104.9&125.3&178.9\\
\hline
{\makecell{{Transformer}}}&187.9&233.6&241.3&377.2\\
\hline
{\makecell{{Bi-Mamba}}}&108.6&156.3&191.8&273.5\\
\hline
{\makecell{{RTSNet}}}&73.4&98.9&118.6&155.9\\
\hline
{\makecell{{EGBRNN}}}&124.1&{171.5}&{205.7}&{266.8}\\
\hline
{\makecell{\textbf{EBRNS}}}&\textbf{61.2}&\textbf{76.5}&\textbf{104.2}&\textbf{129.7}\\
\hline
\end{tabular}
\end{center}
\end{table}

\begin{table}[t]
\renewcommand{\arraystretch}{1.8}
\setlength\tabcolsep{4pt}
\begin{center}
\caption{Mean velocity RMSE (m/s) on the test set.}
\label{Vel_RMSE_tab}
\setlength{\tabcolsep}{.8mm}
\begin{tabular}{ ccccccc }
\hline
 {\makecell{Noise level\\($\sigma_\alpha,\sigma_{\eta}$)}} & {\makecell{$ (0.1^\circ,50\text{m}) $}}&{\makecell{$ (0.15^\circ,100\text{m}) $}}& {\makecell{$ (0.2^\circ,100\text{m}) $}}& {\makecell{$ (0.3^\circ,150\text{m}) $}}\\
\hline
{\makecell{{EKS}}}&12.5&12.8&13.1&13.5\\
\hline
{\makecell{{Transformer}}}&19.0&17.1&16.8&17.3\\
\hline
{\makecell{{Bi-Mamba}}}&48.5&50.4&51.3&51.2\\
\hline
{\makecell{{RTSNet}}}&15.1&15.4&15.8&15.7\\
\hline
{\makecell{{EGBRNN}}}&14.2&{15.1}&{16.5}&{16.9}\\
\hline
{\makecell{\textbf{EBRNS}}}&\textbf{10.4}&\textbf{10.6}&\textbf{11.0}&\textbf{11.5}\\
\hline
\end{tabular}
\end{center}
\end{table}

Tab. \ref{param_RMSE_tab} shows the estimation performance of EBRNS with varying numbers of learnable parameters. Despite the fact that more parameters help it learn the data more adequately, EBRNS maintains good smoothing performance with a limited number of parameters. This is due to the incorporation of the prior model knowledge and the explainable computational architecture, which allow the learnable parameters in EBRNS to have explicit physical meanings. These parameters have their well-defined functions, thus giving EBRNS excellent learning efficiency.

\begin{table}[htb]
\renewcommand{\arraystretch}{2.0}
\setlength\tabcolsep{4pt}
\begin{center}
\caption{Mean position and velocity RMSEs under different number of learnable parameters.}
\label{param_RMSE_tab}
\setlength{\tabcolsep}{.9mm}
\begin{tabular}{ cccccccc }
\hline
{\makecell{Number of\\ learnable parameters}} &{\makecell{$174$}}& {\makecell{$394$}}&{\makecell{$ 1026$}}& {\makecell{$3058$}}& {\makecell{$ 10194$}}& {\makecell{$ 21426$}}& {\makecell{$ 36754$}}\\
\hline
{\makecell{Position RMSE (m)}}&146.2&143.3&135.7&132.1&131.6&129.7&130.0\\
{\makecell{Velocity RMSE (m/s)}}&12.2&12.0&11.9&11.7&11.8&11.5&11.6\\
\hline
\end{tabular}
\end{center}
\end{table}

\section{CONCLUSION}
For the offline data-assisted fixed-interval smoothing problem, we introduce deep learning into Bayesian smoothing through rigorous mathematical derivations under Bayesian estimation theory, proposing an explainable Bayesian recurrent neural smoother, i.e., EBRNS. EBRNS contains two gated recurrent networks for filtering and smoothing, with a bidirectional memory iteration mechanism that captures forward and backward state evolutionary correlations to compensate for uncertain state evolution trends. The bidirectional passing process and internal gated structure of EBRNS are derived from Bayesian estimation theory, which naturally integrates prior model knowledge and offline data with good physical explainability. Experiments on representative real-world datasets demonstrate its smoothing performance, lightweight learnable parameter scale, and low data dependency.

\appendices
\section{Proof of the JSMT-BS}
\label{Proof_A}
From Eq. (\ref{b_mem_iteration}), the backward memory update PDF is $p\left( {{\bf{c}}_k^b\left| {{{\bf{x}}}_{k + 1}},{{\bf{c}}_{k + 1}^b,{\mathcal{D}}} \right.} \right)$. Considering the summarization of state information after time $k$ by the backward memory reflected in Eqs. (\ref{smooth_trans1}) and (\ref{b_error}), based on the Bayes rule, the joint distribution of ${{\bf{x}}_k}$ and ${\bf{c}}_k^b$ conditioned on ${{\bf{x}}_{k + 1}}$, ${\bf{c}}_{k + 1}^b$, ${{\bf{z}}_{1:K}}$ and ${\mathcal{D}}$ can be expressed as
\begin{align}
& p\left( {{{\bf{x}}_k},{\bf{c}}_k^b\left| {{{\bf{x}}_{k + 1}},{\bf{c}}_{k + 1}^b,{{\bf{z}}_{1:K}},{\mathcal{D}}} \right.} \right)
\notag\\
&= p\left( {{\bf{c}}_k^b\left| {{{\bf{x}}_{k + 1}},{\bf{c}}_{k + 1}^b,{\mathcal{D}}} \right.} \right)p\left( {{{\bf{x}}_k}\left| {\bf{c}}_k^b,{{{\bf{x}}_{k + 1}},{\bf{c}}_{k + 1}^b,{{\bf{z}}_{1:k}},{\mathcal{D}}} \right.} \right)
\end{align}

According to Eq. (\ref{b_mem_iteration}), ${\bf{c}}_k^b$ is a function of ${\bf{c}}_{k+1}^b$ and ${\bf{x}}_{k+1}$, it yields
\begin{align}
&p\left( {{{\bf{x}}_k}\left| {\bf{c}}_k^b,{{{\bf{x}}_{k + 1}},{\bf{c}}_{k + 1}^b,{{\bf{z}}_{1:k}},{\mathcal{D}}} \right.} \right) = p\left( {{{\bf{x}}_k}\left|{{{\bf{x}}_{k + 1}},{\bf{c}}_{k + 1}^b,{{\bf{z}}_{1:k}},{\mathcal{D}}} \right.} \right)
\notag \\
&\quad ={{p\left( {{{\bf{x}}_k},{{\bf{x}}_{k + 1}}\left| {{\bf{c}}_{k + 1}^b,{{\bf{z}}_{1:k}},{\cal D}} \right.} \right)} \over {p\left( {{{\bf{x}}_{k + 1}}\left| {{\bf{c}}_{k + 1}^b,{{\bf{z}}_{1:k}},{\cal D}} \right.} \right)}}
\label{PA_1}
\end{align}

For $p\left( {{{\bf{x}}_k},{{\bf{x}}_{k + 1}}\left| {{\bf{c}}_{k + 1}^b,{{\bf{z}}_{1:k}},{\cal D}} \right.} \right)$ in Eq. (\ref{PA_1}), according to the Bayes rule and the model in Eqs. (\ref{smooth_trans1})-(\ref{b_mem_iteration}), we have
\begin{align}
& P_{k + 1}^4 = p\left( {{{\bf{x}}_k},{{\bf{x}}_{k + 1}}\left|{{\bf{c}}_{k + 1}^b,{{\bf{z}}_{1:k}},{\cal D}} \right.} \right) \notag \\
&=p\left( {{{\bf{x}}_{k + 1}}\left| {{{\bf{x}}_k},{\bf{c}}_{k + 1}^b,{{\bf{z}}_{1:k}},{\cal D}} \right.} \right)p\left( {{{\bf{x}}_k}\left|{{{\bf{z}}_{1:k}},{\cal D}} \right.} \right) 
\notag \\
&=\!\! \iiiint \!\! {P_{k + 1}^5 }p\left( {{{\bf{x}}_k},{{\bf{c}}_k^a}\left| {{{\bf{z}}_{1:k}},{\cal D}} \right.} \right) d{\bf \Delta} _{k + 1}^ad{\bf \Delta} _{k + 1}^bd{\bf{c}}_{k+1}^ad{\bf{c}}_k^a
\end{align}
with
\begin{align}
&P_{k + 1}^5  = p\left( {{{\bf{x}}_{k + 1}}\left| {{{\bf{x}}_k},{\bf \Delta} _{k + 1}^a,{\bf \Delta} _{k + 1}^b} \right.} \right)p\left( {{\bf \Delta} _{k + 1}^b\left| {{\bf{c}}_{k + 1}^b},{\cal D} \right.} \right)
\notag \\
&\quad \quad \times
p( {{\bf \Delta} _{k + 1}^a\left| {{\bf{c}}_{k + 1}^a},{\cal D} \right.} )p( {{\bf{c}}_{k + 1}^a\left| {{{\bf{x}}_k},{{\bf{c}}_k^a},{\cal D}} \right.} )
\end{align}

Given ${{\bf{z}}_{1:K}}$ and ${\cal D}$, the joint distribution of ${{\bf{x}}_k}$, ${\bf{c}}_k^b$, ${{\bf{x}}_{k + 1}}$, and ${\bf{c}}_{k + 1}^b$ is
\begin{align}
&p\left( {{{\bf{x}}_k},{\bf{c}}_k^b,{{\bf{x}}_{k + 1}},{\bf{c}}_{k + 1}^b\left| {{{\bf{z}}_{1:K}},{\cal D}} \right.} \right) \notag \\
&= {{p\left( {{\bf{c}}_k^b\left| {{{\bf{x}}_{k + 1}},{\bf{c}}_{k + 1}^b,{\cal D}} \right.} \right)P_{k + 1}^4} \over {p\left( {{{\bf{x}}_{k + 1}}\left| {{\bf{c}}_{k + 1}^b,{{\bf{z}}_{1:k}},{\cal D}} \right.} \right)}}p\left( {{{\bf{x}}_{k + 1}},{\bf{c}}_{k + 1}^b\left| {{{\bf{z}}_{1:K}},{\cal D}} \right.} \right)
\label{joint_proof_1}
\end{align}

By performing an integral operation on Eq. (\ref{joint_proof_1}), we can obtain the JSMT smoothing posterior in Theorem \ref{T}, namely,
\begin{align}
&p\left( {{{\bf{x}}_k},{\bf{c}}_k^b\left| {{{\bf{z}}_{1:K},{\cal D}}} \right.} \right) 
\notag \\
&= \!\!\int\!\!\!\!\int\!\! {p\left( {{{\bf{x}}_k},{\bf{c}}_k^b,{{\bf{x}}_{k + 1}},{\bf{c}}_{k + 1}^b\left| {{{\bf{z}}_{1:K}},{\cal D}} \right.} \right) d{{\bf{x}}_{k + 1}}d{\bf{c}}_{k + 1}^b} 
\end{align}

\section{Proof of the Gaussian approximation implementation for JSMT-BS}
\label{Proof_B}
The proof is based on the computational rules of the Gaussian distribution, which obtains the marginal distribution $p\left( {{{\bf{x}}_k}\left| {{{\bf{z}}_{1:K}},{\cal D}} \right.} \right)$ through the joint distribution $p\left( {{{\bf{x}}_{k + 1}},{{\bf{x}}_k}\left| {{{\bf{z}}_{1:K}},{\cal D}} \right.} \right)$. Firstly, according to Bayes rule, this joint distribution can be expressed as
\begin{align}
&p\left( {{{\bf{x}}_{k + 1}},{{\bf{x}}_k}\left| {{{\bf{z}}_{1:K}},{\cal D}} \right.} \right) 
= \!\!\int\!\! {p\left( {{{\bf{x}}_{k + 1}},{{\bf{x}}_k},{\bf{c}}_{k + 1}^b\left| {{{\bf{z}}_{1:K}},{\cal D}} \right.} \right)} d{\bf{c}}_{k + 1}^b 
\notag \\
&\! = \!\!\!\int\!\! {p\left( {{{\bf{x}}_k}\left| {{{\bf{x}}_{k + 1}},{\bf{c}}_{k + 1}^b,{{\bf{z}}_{1:K}},{\cal D}\!} \right.} \right)\!p\!\left( {{{\bf{x}}_{k + 1}},{\bf{c}}_{k + 1}^b\left| {{{\bf{z}}_{1:K}},{\cal D}} \right.} \right)d{\bf{c}}_{k + 1}^b}
\label{proof_b_1}
\end{align}
with
\begin{align}
&\int {p\left( {{{\bf{x}}_{k + 1}},{\bf{c}}_{k + 1}^b\left| {{{\bf{z}}_{1:K}},{\cal D}} \right.} \right)d} {\bf{c}}_{k + 1}^b = p\left( {{{\bf{x}}_{k + 1}}\left| {{{\bf{z}}_{1:K}},{\cal D}} \right.} \right) 
\notag\\
&= N\left( {{{\bf{x}}_{k + 1}};{{{\bf{\hat x}}}_{k + 1|K}},{{\bf{P}}_{k + 1|K}}} \right)
\end{align}

Next, we compute the Gaussian approximation of $p\left( {{{\bf{x}}_k}\left| {{{\bf{x}}_{k + 1}},{\bf{c}}_{k + 1}^b,{{\bf{z}}_{1:K}},{\cal D}\!} \right.} \right)$. Since the state information at $k+1$ and thereafter is contained in ${\bf{c}}_{k + 1}^b$, it follows that
\begin{align}
&p\left( {{{\bf{x}}_k}\left| {{{\bf{x}}_{k + 1}},{\bf{c}}_{k + 1}^b,{{\bf{z}}_{1:K}},{\cal D}} \right.} \right) =p\left( {{{\bf{x}}_k}\left| {{{\bf{x}}_{k + 1}},{\bf{c}}_{k + 1}^b,{{\bf{z}}_{1:k}},{\cal D}} \right.} \right) 
\end{align}

According to the rules for computing the joint distribution of Gaussian variables (see Lemma A.1 in \cite{Simon_book_2017}), the following joint distribution can be obtained as
\begin{align}
& p\left( {{{\bf{x}}_k},{{\bf{x}}_{k + 1}}\left| {{\bf{c}}_{k + 1}^b,{{\bf{z}}_{1:k}},{\cal D}} \right.} \right) 
\notag \\
& = N\left ( \begin{pmatrix}
 {{{{\bf{\hat x}}}_{k|k}}}\\{{\bf{\hat x}}_{k + 1|k}^b}
\end{pmatrix},\begin{pmatrix}
  {{{\bf{P}}_{k|k}}}&{{{\bf{C}}_{k,k + 1}}} \\
  {{\bf{C}}_{k,k + 1}^{\rm{T}}}&{{\bf{P}}_{k + 1|k}^b}
\end{pmatrix} \right ) 
\label{proof_joint_dis}
\end{align}
with
\begin{align}
&{{{\bf{\hat x}}}_{k + 1|k}}^b = {\rm{E}}\left[ {{{\bf{x}}_{k + 1}}\left|{{\bf{c}}_{k + 1}^b,{{\bf{z}}_{1:k}},{\cal D}} \right.} \right]= {\rm{E}}\left[ {M_{k + 1}^x\left|{{{\bf{c}}_{k + 1}^b,{\bf{z}}_{1:k}},{\cal D}} \right.} \right]
\notag \\
&=\!\!\int\!\!\!\!\int\!\!\!\!\int\!\!\!\!
 \int\!\!\!\!\int \!\!\!\!\
 {M_{k + 1}^xP_k^6d{\bf \Delta} _{k + 1}^ad{\bf \Delta} _{k + 1}^bd{\bf{c}}_{k + 1}^ad{{\bf{x}}_k}d{\bf{c}}_k^a}  
 \label{GA_proof_pred_x}
\end{align}
and
\begin{align}
M_{k + 1}^x = {f_{k + 1}}\left( {{{\bf{x}}_k}} \right) + {\bf \Delta} _{k + 1}^a + {\bf \Delta} _{k + 1}^b + {{\bf{w}}_{k + 1}}
\end{align}
\begin{align}
& P_k^6 = p( {{\bf \Delta} _{k + 1}^a\left| {{\bf{c}}_{k + 1}^a},{\cal D} \right.} )p( {{\bf \Delta} _{k + 1}^b\left| {{\bf{c}}_{k + 1}^b},{\cal D} \right.} )\notag \\
& \quad \ \times p( {{\bf{c}}_{k + 1}^a\left| {{{\bf{x}}_k},{\bf{c}}_k^a},{\cal D} \right.} )p( {{{\bf{x}}_k},{\bf{c}}_k^a\left| {{{\bf{z}}_{1:k}},{\cal D}} \right.} )
\end{align}

Substituting the corresponding Gaussian assumptions into Eq. (\ref{GA_proof_pred_x}), we can get the computation of ${{{{\bf{\hat x}}}}_{k + 1|k}}^b$ in Eq. (\ref{Back_state_pred}).

Let $V_{k + 1}^x =  {M_{k + 1}^x - {\bf{\hat x}}_{k + 1|k}^b} $, then we have
\begin{align}
&{\bf{P}}_{k + 1|k}^b = {\rm{E}}\left[ {\left( {{{\bf{x}}_{k + 1}} - {\bf{\hat x}}_{k + 1|k}^b} \right){{\left(  \cdot  \right)}^{\rm{T}}}\left| {{\bf{c}}_{k + 1}^b,{{\bf{z}}_{1:k}},{\cal D}} \right.} \right] 
\notag \\
&= \!\!\int\!\!\!\!\int\!\!\!\!\int\!\!\!\!
 \int\!\!\!\!\int \!\!{{{V_{k + 1}^x}({V_{k + 1}^x})^{\rm{T}} } } P_k^6d{\bf \Delta} _{k + 1}^ad{\bf \Delta} _{k + 1}^bd{\bf{c}}_{k + 1}^ad{{\bf{x}}_k}d{\bf{c}}_k^a
\end{align}

Similarly, substituting the corresponding Gaussian assumptions, we obtain the computation of ${\bf{P}}_{k + 1|k}^b$ as shown in (\ref{Back_cov_pred}).

Let $P_k^{x + }=N\left( {{{\bf{x}}_{k|k}};{{{\bf{\hat x}}}_{k|k}},{{\bf{P}}_{k|k}}} \right)$, and consider the corresponding Gaussian assumptions, we have
\begin{align}
&{{\bf{C}}_{k,k + 1}} = {\rm{E}}\left[ {( {{{\bf{x}}_k} - {{{\bf{\hat x}}}_{k|k}}} ){{( {{{\bf{x}}_{k + 1}} - {\bf{\hat x}}_{k + 1|k}^b} )}^{\rm{T}}}\left| {{\bf{c}}_{k + 1}^b,{{\bf{z}}_{1:k}},{\cal D}} \right.} \right]
\notag \\
&  = \int\!\! {\left( {{{\bf{x}}_k} - {{{\bf{\hat x}}}_{k|k}}} \right)} {\left( {f_{k+1}\left( {{{\bf{x}}_k}} \right) - f_{k+1}\left( {{{{\bf{\hat x}}}_{k|k}}} \right) + {{\bf{w}}_k}} \right)^{\rm{T}}}P_k^{x + }d{{\bf{x}}_k}
\notag \\
&= \int\!\! {{{\bf{x}}_k}{{\left( {f_{k+1}\left( {{{\bf{x}}_k}} \right)} \right)}^{\rm{T}}} } P_k^{x + }d{{\bf{x}}_k} + {{\bf{\hat x}}_{k|k}}{\left( {f_{k+1}\left( {{{{\bf{\hat x}}}_{k|k}}} \right)} \right)^{\rm{T}}}
\end{align}

Let $p\left( {{{\bf{x}}_k}\left| {{{\bf{x}}_{k + 1}},{\bf{c}}_{k + 1}^b,{{\bf{z}}_{1:k}},{\cal D}} \right.} \right) = N\left( {{{\bf{x}}_k};{\bf{x}}_{k|k}^m,{\bf{P}}_{k|k}^m} \right)$. According to the conditional distribution rule for Gaussian variables (see Lemma A.2 in \cite{Simon_book_2017}), based on Eq. (\ref{proof_joint_dis}), we have
\begin{align}
&{\bf{x}}_{k|k}^m = {{\bf{x}}_{k|k}} + {{\bf{G}}_k}\left( {{{\bf{x}}_{k + 1}} - {\bf{\hat x}}_{k + 1|k}^b} \right)
\\
&{\bf{P}}_{k|k}^m = {{\bf{P}}_{k|k}} - {{\bf{G}}_k}{\bf{P}}_{k + 1|k}^b{\bf{G}}_k^{\rm{T}}
\end{align}
with the smoothing gain is
\begin{align}
{{\bf{G}}_k} = {{\bf{C}}_{k,k + 1}}{\left( {{\bf{P}}_{k + 1|k}^b} \right)^{ - 1}}
\end{align}

From Eq. (\ref{proof_b_1}), the joint distribution of ${{\bf{x}}_{k + 1}}$ and ${{\bf{x}}_{k}}$ for given all measurements is
\begin{align}
p\left( {{{\bf{x}}_{k + 1}},{{\bf{x}}_k}\left| {{{\bf{z}}_{1:K}},{\cal D}} \right.} \right)
=N\left ( \begin{pmatrix}
 {{{{\bf{x}}}_{k+1}}}\\{{\bf{x}}_{k}}
\end{pmatrix};a_k,A_k \right ) 
\end{align}
with
\begin{align}
&a_k=\begin{pmatrix}
 {{{{\bf{\hat x}}}_{k + 1|K}}}\\{{{{\bf{\hat x}}}_{k|k}} + {{\bf{G}}_k}( {{{{\bf{\hat x}}}_{k + 1|K}} - {\bf{\hat x}}_{k + 1|k}^b} )}
\end{pmatrix}
\\
&A_k=\begin{pmatrix}
  {{{\bf{P}}_{k + 1|K}}}&{{{\bf{P}}_{k + 1|K}}{\bf{G}}_k^{\rm{T}}} \\
  {{{\bf{G}}_k}{{\bf{P}}_{k + 1|K}}}&{{{\bf{P}}_{k|k}} + {{\bf{G}}_k}\left( {{{\bf{P}}_{k + 1|K}} - {\bf{P}}_{k + 1|k}^b} \right){\bf{G}}_k^{\rm{T}}}
  \end{pmatrix}
\end{align}

Finally, according to the conditional distribution rule for the Gaussian variable, we can get the marginal distribution $p\left( {{{\bf{x}}_k}\left| {{{\bf{z}}_{1:K}},{\cal D}} \right.} \right)$ as in Eqs. (\ref{smooth_x}) and (\ref{smooth_P}).

\bibliographystyle{IEEEtran}
\bibliography{ref}


	



\end{document}